\documentclass[aps,pra,showpacs,floatfix,twocolumn]{revtex4}
\usepackage{graphicx} % Include figure files
\usepackage{amsmath}
\usepackage{bm}
\usepackage{dcolumn}% Align table columns on decimal point
\begin{document}

\title{Blackbody radiation shift, multipole polarizabilities, oscillator strengths, lifetimes, hyperfine constants,
 and excitation  energies  in Ca$^+$ }

\author{ M. S. Safronova$^1$ and  U. I. Safronova$^{2,3}$ }

\affiliation {$^1$Department of Physics and Astronomy, 217 Sharp
Lab, University of Delaware, Newark, Delaware 19716\\
$^2$Physics Department, University of Nevada, Reno, Nevada 89557,
\\$^3$Institute of Spectroscopy, Russian Academy of Science,
Troitsk, Moscow, Russia}

\date{\today}
\begin{abstract}
 A systematic study of Ca$^+$ atomic properties is carried out using high-precision  relativistic all-order
 method where all single, double, and partial triple excitations
of the Dirac-Fock wave functions are included to all orders of perturbation theory. Reduced matrix elements, oscillator strengths, transition rates,
and lifetimes are determined for the levels up to $n = 7$. Recommended values and estimates of their uncertainties are provided for a large number of
electric-dipole transitions. Electric-dipole scalar polarizabilities for the $5s$, $6s$, $7s$, $8s$, $4p_j$, $5p_j$, $3d_j$, and $4d_j$  states and
tensor polarizabilities for the $4p_{3/2}$, $5p_{3/2}$, $3d_j$, and $4d_j$ states in Ca$^+$ are calculated. Methods are developed to accurately treat
the contributions from highly-excited states, resulting in significant (factor of 3) improvement in accuracy of the $3d_{5/2}$ static  polarizability
value, $31.8(3)~a^3_0$, in comparison with the previous calculation [Arora \textit{et al.}, Phys. Rev. A 76, 064501 (2007)]. The blackbody radiation
(BBR) shift of the $4s - 3d_{5/2}$ clock transition in Ca$^+$
 is calculated to be $0.381(4)$~Hz at room temperature, $T=300$~K.
Electric-quadrupole $4s -nd$ and electric-octupole $4s -nf$ matrix elements are calculated to obtain the ground state multipole E2 and E3 static
polarizabilities.
 Excitation energies  of the  $ns$,  $np$, $nd$, $nf$, and $ng$ states with
$n \leq$ 7 in are evaluated and compared with experiment. Recommended values are provided for the $7p_{1/2}$, $7p_{3/2}$,
 $8p_{1/2}$, and $8p_{3/2}$ removal energies for which experimental measurements are not available.
 The hyperfine constants $A$ are determined for the low-lying levels up to $n$ = 7. The quadratic Stark
effect on hyperfine structure levels of $^{43}$Ca$^+$ ground state is investigated.  These calculations provide recommended values critically
evaluated for their accuracy for a number of Ca$^+$ atomic properties for use in planning and analysis of various experiments as well as theoretical
modeling.

%31.15.A- Ab initio calculations
%31.15.ac High-precision calculations for few-electron (or few-body) atomic systems
%31.15.ag Excitation energies and lifetimes; oscillator strengths
%31.15.aj Relativistic corrections, spin-orbit effects, fine structure; hyperfine structure
%31.15.am Relativistic configuration interaction (CI) and many-body perturbation calculations
%31.15.ap Polarizabilities and other atomic and molecular properties
%31.15.V- Electron correlation calculations for atoms, ions and molecules
%31.15.vj Electron correlation calculations for atoms and ions: excited states
%31.30.J- Relativistic and quantum electrodynamic (QED) effects in atoms, molecules, and ions
\end{abstract}
% It is always \today, today,
%  but any date may be explicitly specified
% PACS, the Physics and Astronomy
% Classification Scheme.
%\keywords{Suggested keywords}%Use showkeys class option if keyword
%display desired
\maketitle

%*************************
\section{Introduction}

This work presents a systematic study of atomic properties of Ca$^+$ ion motivated by its importance for the development of optical frequency
standards and quantum information processing.

The current definition of a second in the International System of Units (SI)  is based on the microwave transition between the two hyperfine levels
of the ground state of $^{133}$Cs. The present relative standard uncertainty of Cs microwave frequency standard is around $4\times10^{-16}$
\cite{nist-cs}. More precise frequency standards will open ways to more sensitive quantum-based standards for applications such as measurements of
the fundamental constants and testing of physics postulates, inertial navigation, magnetometry, gravity gradiometry, and tracking of deep-space
probes. Optical frequency standards may achieve even smaller relative uncertainties owing to superior resonance line quality factors, allowing
shorter averaging times and higher stability. Significant recent progress in optical spectroscopy and
 measurement techniques has led to the achievement of relative standard uncertainties in

  optical frequency standards that are comparable to the Cs microwave benchmark.
With extremely low systematic perturbations and better stability and accuracy, such optical frequency standards can
reach a systematic fractional uncertainty of the order of $10^{-18}$~\cite{Margolis,uncertainty}.

Prospects of optical frequency standard based on the metastable $4s-3d_{5/2}$ transition in Ca$^+$ ion have been studied in
\cite{clock2,clock3,clock4}. In 2009, the first absolute transition frequency measurement at the $10^{-15}$ level with a single, lasercooled
$^{40}$Ca$^+$ ion in a linear Paul trap has been reported \cite{clock1}. The development of an ion clock based on Ca$^+$ has the technological
advantage that all necessary wavelengths for laser cooling and state manipulation including lasers for photoionization can be generated by
commercially available and easy-to-handle solid state lasers ~\cite{clock1,clock3}. The operation of atomic clocks is generally carried out at room
temperature, whereas the definition of the second refers to the clock transition in an atom at absolute zero. This implies that the clock transition
frequency should be corrected in practice for the effect of finite temperature of which the leading contributor is the blackbody radiation (BBR)
shift. Recent experimental work~\cite{clock3} notes that uncertainty due to BBR shift is particularly difficult to improve by experimental means.
This uncertainty results both from the uncertainty in the stability and accuracy of trap temperature measurement as well as the uncertainty in the
evaluation of the BBR shift coefficient (i.e. BBR shift at 300K). In the present paper, we improve the accuracy of the BBR shift value at 300K by a
factor of 3.

Ca$^+$ ions have been used for a number of quantum information processing experiments (see Refs.~\cite{quantum1,quantum2} and references therein).
Most of the elementary building blocks for quantum information processing such as state initialization, long quantum information storage times,
universal set of quantum logic gates, and readout have been realized with high fidelity with trapped ion systems such as Ca$^+$
\cite{1a,2a,3a,4a,5a,6a}. Recently, the merits of a high-fidelity entangling operation on an optical transition (optical qubit) were combined with
the long coherence times of two "clock" states in the hyperfine ground state (hyperfine qubit) by mapping between these two qubits \cite{quantum2}.
Precise understanding of the ion qubit and gate operation decoherence properties is aided by precise knowledge of atomic properties of this system.

Properties of Ca$^+$ are also of interest to astrophysics as the absorption spectrum of the Ca$^+$ ion is used to explore the structure and
properties of interstellar dust clouds \cite{1b,2b}. Below, we briefly review previous studies of Ca$^+$ atomic properties.

 The lifetime of the metastable  $3d_{3/2}$ and $3d_{5/2}$ levels in Ca$^+$ was a subject of many theoretical
and experimental studies owing to importance of these states for the
  optical frequency standards and quantum information applications.
Early theoretical calculations and measurements of the $3d_{3/2}$ and $3d_{5/2}$ lifetimes in Ca$^+$ were reviewed in Ref.~\cite{ca-3d-expt-05}. Both
high precision measurements and calculations of the $3d_{j}$ lifetimes were presented by Kreuter {\it et al.\/} in Ref.~\cite{ca-3d-expt-05}. A
measurement technique was based on high-efficiency quantum state detection after coherent excitation to the $3d_{5/2}$ state or incoherent shelving
in the $3d_{3/2}$ state, and subsequent free, unperturbed spontaneous decay. The calculation of the $3d_j-4s$ electric-quadrupole matrix elements was
carried out using an {\it ab initio} relativistic all-order method which sums infinite sets of many-body perturbation theory terms. These matrix
elements were used to evaluate the $3d$ radiative lifetimes and their ratio \cite{ca-3d-expt-05}. In Refs.~\cite{prl-das-06,nd-das-06}, the
relativistic coupled-cluster theory was used to perform the  calculations of these lowest excited $3d_{3/2}$ and $3d_{5/2}$ state lifetimes.

 The
blackbody radiation (BBR) shift of the $4s - 3d_{5/2}$ clock transition, accurate to 3\%,  and the $3d_{5/2}$ tensor polarizability were  presented
by Arora {\it et al.\/} in Ref.~\cite{mar-bbr-07}. The calculations were based on the relativistic all-order single-double method where all single
and double excitations of the Dirac-Fock wave function are included to all orders of perturbation theory. The largest contribution to the uncertainty
of the BBR shift originated from the contribution of the highly-excited $nf_{7/2}$ states to the $3d_{5/2}$ static polarizability.

Relativistic coupled-cluster studies of ionization potentials, lifetimes, and polarizabilities in singly ionized calcium was recently presented by
Sahoo {\it et al.\/} in Ref.~\cite{das-09}. Numerical results were given  for the $4p_j$ lifetime and the $4s$ and $3d_j$ polarizabilities. The
polarizabilities of the $4s$, $5s$, $4p$, $5p$, 3d, $4d$ of the Ca$^+$ ions were recently evaluated  by Mitroy and Zhang \cite{mitroy-08} using the
non-relativistic  configuration interaction with semi-empirical core potential (CICP) method .

 The first  measurement of the $4p_{3/2}$ lifetime  in Ca$^+$ was presented by Smith and Gallagher \cite{life-4p-expt-66}
 using Hanle-effect method with optical excitation from the $4s$ ground state. The same technique
 was used by Gallagher \cite{osc-expt-67} to measure  the branching ratio of the $4p_{3/2}$ decay
[17.6(2.0)]. The beam-foil technique was used by Andersen {\it et al.\/}~\cite{life-expt-70} to measure the oscillator strengths for the  $4s-4p$,
$4p-5s$, $4p-4d$, and $4p-5d$ transitions. The same technique was used by Emmoth {\it et al.\/}~\cite{life-expt-75}. Additionally, the effects of
cascades were analyzed and corrected for lifetime measurements. The first pulsed laser excitation measurements of the $4p_{1/2}$ and $4p_{3/2}$ level
lifetime (6.96$\pm$0.35~ns and 6.71$\pm$0.25~ns, respectively) were reported by Ansbacher {\it et al.\/}~\cite{life-4p-expt-85}. Gosselin {\it et
al.\/} presented precision $4p_{1/2}$ and $4p_{3/2}$ lifetime measurements in Refs.~\cite{life-4p-expt-88-nim,life-4p-expt-88-pra}. Two sets of
results (6.95$\pm$0.18~ns and 6.87$\pm$0.17~ns \cite{life-4p-expt-88-nim}) and (7.07$\pm$0.07~ns and 6.87$\pm$0.06~ns \cite{life-4p-expt-88-pra})
agree within their uncertainties. The mean lifetimes of the $4p_{1/2}$ and $4p_{3/2}$ levels in Ca$^+$ were measured by Jin and Church
\cite{life-4p-expt-93,life-4p-expt-94} to 0.3\% precision using a variant of the collinear laser-beam-ion-beam spectroscopy technique
(7.098$\pm$0.020~ns and 6.924$\pm$0.019~ns, respectively). Lifetime  of the $4p_{3/2}$ level (6.94$\pm$0.18~ns) was measured by Rosner {\it et
al.\/}~\cite{life-4p-expt-97} using  the cascade-photon-coincidence technique with a sputtered-atom source. We note that recent linearized
coupled-cluster calculation \cite{mar-bbr-07}  disagrees with 0.3\% Jin and Church measurement by 3\%. The accuracy of the calculations should be
better than 1\% owing to excellent agreement of similar calculations for all alkali-metal atoms from Li to Fr \cite{mar-pol-99}, Mg$^+$\cite{mg},
Sr$^+$\cite{sr}, and Ba$^+$\cite{ba} with all recent experimental values.

Precision measurement of the branching ratios of the $4p_{3/2}$ decay of Ca$^+$ was performed recently by Gerritsma {\it et al.\/} in
Ref.~\cite{branch-08}. High precision was achieved by a novel technique based on monitoring the population transfer when repeatedly pumping the ion
between different internal states. Authors underlined that forty-fold improvement was achieved for the $A(4p_{3/2}-4s)$/$\sum_{J} A(4p_{3/2}-nd_{j})$
= 14.31(5) branching ratio in  comparison with the best previous measurement \cite{branch-08}.

Warner reported \cite{e2-68} oscillator strengths for the $ns-n'p$, $np-n'd$, and $nd-n'f$ transitions with $ns = 4s - 9s$, $np = 4p - 9p$, $nd = 3d
- 7d$, and $nf = 4f - 7f$.
 The radial wave functions were calculated
using scaled Thomas-Fermi-Dirac wave functions with including spin-orbit interaction \cite{e2-68}. The relativistic pseudopotential approach was
applied by Hafner and Schwarzt \cite{tran-jpb-75} to the calculation of the $ns-n'p$ electric-dipole transition probabilities for $n = 4-7$ and $n' =
4 - 6$.  Semiemerical wave functions were used by Theodosiou \cite{life-th-89} to evaluate oscillator strengths and lifetimes of the $5s$, $6s$,
$7s$, $4p$, $5p$, $6p$, $4d$, $5d$, and $4f$ levels. Relativistic many-body theory was applied by Guet and Johnson \cite{nd-wrj-91} to determine
amplitudes for the $4p-4s$ and $4p-3d$ transitions and the $4p$ lifetime. Multi-configuration Hartree-Fock wave functions were used by Vaeck {\it et
al.\/}~\cite{ca-pra-92-froese} to evaluate oscillator strengths for the $4p-4s$ and $4p-3d$ transitions and the $4p$ lifetime  in Ca$^+$. Liaw
\cite{ca-liaw-95} performed {\it ab initio} calculations based on the Brueckner approximation for the amplitudes of the $4p-4s$ and $4p-3d$
transitions and the $4p$ lifetime in Ca$^+$. The lifetimes of the $5s$, $6s$, $4p$, $5p$, $6p$, $4d$,  $5d$, and $4f$ levels in Ca$^+$ were evaluated
by Mel\'{e}ndez {\it et al.\/}~\cite{life-3d-th-07} using the Thomas-Fermi-Dirac central potential method in the frozen core approximation and
including the polarization interaction between the valence electron and the core using a model potential. Recently, non-relativistic CICP method was
used by Mitroy {\it et al.\/}~\cite{osc-mitroy-88} to evaluate the $4s-np$, $4p-ns$, $3d-np$ oscillator strengths with $n = 4$, $5$.

The hyperfine structure of the $4s$, $4p$, and $3d$ states was evaluated by M\aa rtensson-Pendrill and Salomonson \cite{hyp-84} using many-body
perturbation theory. Numerical values were given for  the $A(4s)$, $A(4p)$, and $A(3d)$ magnetic-dipole hyperfine constants and for the $B(4p_{3/2})$
and $B(3d)$ electric-quadrupole hyperfine constants \cite{hyp-84}. First measurements published several years later \cite{hyp-expt-90} confirmed
theoretical predictions \cite{hyp-84}. The experimental and theoretical results for the $A(4s_{1/2})$ and $A(4p_{j})$ hyperfine constants agreed at
the 3\% level \cite{hyp-expt-90}. Additional correlation contributions were added to the  method used in \cite{hyp-84} by M\aa rtensson-Pendrill {\it
et al.\/}~\cite{hyp-th-92} to recalculate all of the above magnetic-dipole hyperfine constants. The hyperfine structure splittings of the $4s$,
$4p_{1/2}$, and $4p_{3/2}$ levels in $^{43}$Ca~II were measured by fast ion beam collinear laser spectroscopy in \cite{hyp-91Z}. Precise
determination of the $4s$ ground state hyperfine structure splitting of $^{43}$Ca$^+$ (3225.6082864(3)MHz) was reported by Arbes {\it et
al.\/}~\cite{hyp-94Z}. The Doppler-free and potentially very narrow resonances were used in Ref.~\cite{hyp-95Z} to determine the magnetic dipole
hyperfine interaction constant $A$ for the $4p_{1/2}$ and $3d_{3/2}$ states of $^{43}$Ca$^+$. Hyperfine structure in the three
$3d_{3/2,5/2}\Rightarrow 4p_{1/2,3/2}$  transitions  were studied by fast ion beam collinear laser spectroscopy for all stable Ca isotopes in
\cite{hyp-expt-98}. Hyperfine structure parameters $A(4p)$,  $A(3d)$, $B(4p_{3/2})$, and $B(3d_{j})$ for the odd isotope $^{43}$Ca$^+$, as evaluated
from the splittings observed, agreed well with theoretical predictions from relativistic many-body perturbation theory \cite{hyp-84,hyp-th-92}.
Recently, relativistic many-body calculations were performed to calculate the magnetic-dipole hyperfine constants $A(4s)$, $A(4p)$, and $A(3d)$ and
the electric quadrupole constants $B(4p_{3/2})$ and $B(3d)$ in $^{43}$Ca~II~\cite{hyp-das-03,hyp-04}. The relativistic coupled cluster theory was
employed by Sahoo {\it et al.\/}~\cite{hyp-das-03} to calculate the hyperfine $A$ constants of the low-lying states.

\begin{table*}
\caption{\label{tab1} Zeroth-order (DF), second-, and third-order Coulomb correlation energies $E^{(n)}$,
single-double Coulomb energies $E^\text{{SD}}$, $E^{(3)}_\text{{extra}}$,
 first-order Breit and second-order Coulomb-Breit
 corrections $B^{(n)}$  to the energies of Ca~II.
The total energies ($E^{(3)}_\text{ tot} = E^{(0)} + E^{(2)} +E^{(3)}+ B^{(1)} + B^{(2)} + E^{\rm (LS)}$, $E^\text{SD}_\text{tot} = E^{(0)} +
E^\text{{SD}} + E^{(3)}_\text{{extra}} + B^{(1)} + B^{(2)} + E^{\rm (LS)}$) of  Ca~II are compared with experimental
 energies $E_\text{{NIST}}$ \protect\cite{nist-web},
 $\delta E$ = $E_\text{tot}$ - $E_\text{{NIST}}$. Units: cm$^{-1}$. $^a$ Recommended values.}
\begin{ruledtabular}\begin{tabular}{lrrrrrrrrrcrrr}
\multicolumn{1}{c}{$nlj$ } & \multicolumn{1}{c}{$E^{(0)}$} & \multicolumn{1}{c}{$E^{(2)}$} &
\multicolumn{1}{c}{$E^{(3)}$} & \multicolumn{1}{c}{$B^{(1)}$} & \multicolumn{1}{c}{$B^{(2)}$} &
\multicolumn{1}{c}{$E^{\rm (LS)}$} & \multicolumn{1}{c}{$E^{(3)}_\text{ tot}$} & \multicolumn{1}{c}{$E^\text{{SD}}$} &
\multicolumn{1}{c}{$E^{(3)}_\text{{extra}}$} & \multicolumn{1}{c}{$E^\text{{SD}}_\text{ tot}$} &
\multicolumn{1}{c}{$E_\text{{NIST}}$} & \multicolumn{1}{c}{$\delta E^{(3)}$} &
\multicolumn{1}{c}{$\delta E^\text{{SD}}$} \\
\hline
$4s_{1/2}$&  -91440&  -4786.3&     857.6&   23.8&  -23.2&   3.5& -95364&   -4697.6 &    520.2&   -95600&  -95752&    387&   139\\
$4p_{1/2}$&  -68037&  -2675.3&     411.7&   19.7&  -12.3&   0.0& -70293&   -2752.7 &    271.8&   -70506&  -70560&    267&    50\\
$4p_{3/2}$&  -67837&  -2642.5&     406.0&   14.2&  -13.3&   0.0& -70073&   -2719.4 &    268.2&   -70283&  -70337&    265&    50\\
$3d_{3/2}$&  -72617& -10333.3&    1989.7&   34.1&  -99.8&   0.0& -81027&  -10578.5 &   1010.7&   -82206&  -82102&   1075&  -149\\
$3d_{5/2}$&  -72593& -10277.7&    1978.0&   21.0&  -97.8&   0.0& -80970&  -10520.6 &   1003.4&   -82142&  -82041&   1071&  -146\\
$4d_{3/2}$&  -37034&  -2209.5&     395.1&    8.0&  -21.7&   0.0& -38862&   -1938.7 &    216.3&   -38761&  -38913&     51&   143\\
$4d_{5/2}$&  -37018&  -2200.9&     393.5&    5.0&  -21.4&   0.0& -38841&   -1932.8 &    215.1&   -38743&  -38893&     52&   142\\
$4f_{5/2}$&  -27473&   -224.8&      22.2&    0.0&   -0.1&   0.0& -27676&    -237.6 &     22.8&   -27688&  -27695&     19&     7\\
$4f_{7/2}$&  -27473&   -224.7&      22.2&    0.0&   -0.1&   0.0& -27676&    -237.6 &     22.8&   -27688&  -27695&     19&     7\\
$5s_{1/2}$&  -42428&  -1314.5&     242.8&    7.4&   -7.0&   0.5& -43499&   -1246.8 &    145.2&   -43525&  -43585&     86&    56\\
$5p_{1/2}$&  -34406&   -877.1&     140.2&    7.3&   -4.5&   0.0& -35140&    -868.4 &     90.0&   -35180&  -35219&     79&    37\\
$5p_{3/2}$&  -34333&   -868.4&     138.6&    5.3&   -4.9&   0.0& -35062&    -860.1 &     89.0&   -35102&  -35141&     78&    37\\
$5d_{3/2}$&  -22244&   -910.4&     156.6&    3.3&   -8.8&   0.0& -23003&    -809.7 &     89.2&   -22967&  -23030&     26&    60\\
$5d_{5/2}$&  -22236&   -907.7&     156.1&    2.1&   -8.7&   0.0& -22994&    -807.8 &     88.8&   -22958&  -23021&     27&    59\\
$5f_{5/2}$&  -17589&   -129.7&      13.4&    0.0&   -0.1&   0.0& -17706&    -137.2 &     13.1&   -17713&  -17717&     12&     4\\
$5f_{7/2}$&  -17589&   -129.6&      13.3&    0.0&   -0.1&   0.0& -17706&    -137.2 &     13.0&   -17713&  -17717&     12&     4\\
$6s_{1/2}$&  -24589&   -556.3&     103.9&    3.3&   -3.1&   0.1& -25041&    -521.3 &     61.7&   -25046&  -25074&     34&    26\\
$6p_{1/2}$&  -20894&   -405.0&      65.4&    3.5&   -2.2&   0.0& -21232&    -401.5 &     41.7&   -21252&  -21267&     35&    15\\
$6p_{3/2}$&  -20859&   -401.4&      64.8&    2.5&   -2.3&   0.0& -21195&    -398.0 &     41.3&   -21215&  -21230&     35&    15\\
$6d_{3/2}$&  -14820&   -472.5&      79.6&    1.7&   -4.5&   0.0& -15215&    -423.2 &     46.3&   -15198&  -15230&     15&    31\\
$6d_{5/2}$&  -14815&   -471.2&      79.4&    1.1&   -4.5&   0.0& -15211&    -422.4 &     46.1&   -15193&  -15226&     15&    31\\
$6f_{5/2}$&  -12215&    -79.6&       8.3&    0.0&    0.0&   0.0& -12286&     -84.2 &      8.0&   -12291&  -12294&      7&     2\\
$6f_{7/2}$&  -12215&    -79.6&       8.3&    0.0&    0.0&   0.0& -12286&     -84.2 &      8.0&   -12291&  -12294&      7&     2\\
$7s_{1/2}$&  -16053&   -288.7&      54.1&    1.7&   -1.6&   0.0& -16287&    -268.8 &     32.1&   -16289&  -16304&     16&    14\\
$7p_{1/2}$&  -14051&   -221.0&      35.8&    1.9&   -1.2&   0.0& -14236&    -218.3 &     22.8&   -14246&        &       &      \\
          &          &         &        &         &       &      &     &           &         &   -14253$^a$&       &  & \\
$7p_{3/2}$&  -14032&   -219.2&      35.5&    1.4&   -1.3&   0.0& -14215&    -216.6 &     22.6&   -14225&        &       &       \\
         &          &         &        &         &       &      &     &           &          &   -14232$^a$&       &  & \\
$7d_{3/2}$&  -10576&   -278.3&      46.3&    1.0&   -2.6&   0.0& -10809&    -250.2 &     27.3&   -10799&  -10818&      9&    18\\
$7d_{5/2}$&  -10573&   -277.6&      46.2&    0.6&   -2.6&   0.0& -10806&    -249.7 &     27.2&   -10796&  -10815&      9&    18\\
$7f_{5/2}$&   -8974&    -51.8&       5.5&    0.0&    0.0&   0.0&  -9020&     -54.9 &      5.2&    -9023&   -9025&      5&     2\\
$7f_{7/2}$&   -8974&    -51.8&       5.5&    0.0&    0.0&   0.0&  -9020&     -54.8 &      5.2&    -9023&   -9025&      5&     1\\
 \end{tabular}
\end{ruledtabular}
\end{table*}
Despite many previous studies, no reliable recommended values exist for a number of properties of low-lying states of Ca$^+$. In many cases, only
semi-empirical calculations are available. In this work, we carry out a systematic study of Ca$^+$ energies, E1, E2, E3 matrix elements, transition
rates, lifetimes, $A$ and $B$ hyperfine constants, E2 and E3 ground state polarizabilities, scalar E1 polarizabilities of the $5s$, $6s$, $7s$, $8s$,
$4p_j$, $5p_j$, $3d_j$, $4d_j$ states, and tensor polarizabilities of the  $4p_{3/2}$, $5p_{3/2}$, $3d_j$, and $4d_j$ states using a high-precision
all-order approach~\cite{review07} in several different approximations. We evaluate the uncertainties of our calculations for most of the values
listed in this work. We also re-evaluated the blackbody-radiation shift in a Ca$^+$ ion optical frequency standard and improved its accuracy by a
factor of 3. The quadratic Stark effect on the hyperfine structure levels of $^{43}$Ca$^+$ ground state is investigated. The methodologies for
evaluating the uncertainties of theoretical values calculated in the framework of the all-order approach are developed. The calculation of
uncertainties involved estimation of missing high-order effects and {\it ab initio} calculations in different approximations to establish the size of
the higher-order corrections and approximate missing contributions. We evaluated the uncertainties of the recommended values for the transition
matrix elements, oscillator strengths, transition rates, lifetimes, polarizabilities, BBR shift, and the Stark shift coefficient.

The main motivation for this work is to provide recommended values critically evaluated for their accuracy  for a number of atomic properties via a
systematic high-precision study for use in planning and analysis of various experiments as well as theoretical modeling.

%*******************
\section{Energy levels}

Energies of the $nl_j$ states in Ca~II are evaluated for $n\leq 7$ and $l\leq 3$ using both third-order relativistic many-body perturbation theory
(RMBPT) and the single-double (SD) all-order method. The all-order (linearized coupled-cluster) method and its applications are discussed in detail
in review~\cite{review07} and references therein. Therefore, we do no repeat the method description in this work, with the exception of the details
needed to discuss the evaluation of uncertainties in the transition matrix elements carried out in the next section. The comparison of the calculated
energy levels with the experimental values gives an excellent indication of the accuracy of the approach and identifies state with particularly large
correlation corrections. Moreover, a number of highly-excited energy levels that we need for the polarizability calculations appear to be not known
and are missing from the NIST database \cite{nist-web}. Results of our energy calculations are summarized in Table~\ref{tab1}. Columns 2--7 of
Table~\ref{tab1} give the lowest-order DF energies $E^{(0)}$, second-order and third-order Coulomb correlation energies $E^{(2)}$ and $E^{(3)}$,
first-order and second-order Breit corrections $B^{(1)}$ and $B^{(2)}$, and an estimated Lamb shift contribution, $E^{\rm (LS)}$. The Lamb shift
$E^{\rm (LS)}$ is calculated as the sum of the one-electron self energy and the first-order vacuum-polarization energy. The vacuum-polarization
contribution is calculated from the Uehling potential using the results of Fullerton and Rinker \cite{vacuum}. The self-energy contribution is
estimated for the $s$, $p_{1/2}$ and $p_{3/2}$ orbitals by interpolating among the values obtained by \citet{mohr1,mohr2,mohr3} using Coulomb wave
functions. For this purpose, an effective nuclear charge $Z_\text{eff}$ is obtained by finding the value of $Z_\text{eff}$ required to give a Coulomb
orbital with the same average $\langle r\rangle$ as the DF orbital. We find that the values of $E^{\rm (LS)}$ are very small.
 For states with
$l>0$, the Lamb-shift is estimated to be smaller than 0.1 cm$^{-1}$ using scaled Coulomb values and is negligible at the present level of accuracy.
We list the all-order SD energies in the column labeled $E^\text{SD}$ and the part of the third-order energies missing from $E^\text{SD}$ in the
column labeled $E^{(3)}_\text{{extra}}$. The sum of the seven terms $E^{(0)}$, $E^\text{SD}$, $E^{(3)}_\text{{extra}}$, $B^{(1)}$, $B^{(2)}$, and
$E^{\rm (LS)}$
 gives our final all-order result $E^\text{SD}_\text{tot}$, listed
in the eleventh column of Table~\ref{tab1}.  Recommended energies from the National Institute of Standards and
Technology (NIST) database \cite{nist-web} are given in the column labeled $E_\text{NIST}$. Differences between our
third-order and all-order calculations and experimental data, $\delta E^{(3)}=E^{(3)}_{\rm tot}-E_\text{NIST}$ and
$\delta E^\text{SD}=E^\text{SD}_{\rm tot}-E_\text{NIST}$, are given in the two final columns of Table~\ref{tab1},
respectively.

As expected, the largest correlation contribution to the valence energy  comes from the second-order term $E^{(2)}$.
Therefore, we calculate $E^{(2)}$  with higher numerical accuracy. The second-order energy includes partial waves up to
$l_{\text{max}}=8$ and is extrapolated to account for contributions from higher partial waves (see, for example,
Refs.~\cite{be-en,be3-en} for details of the extrapolation procedure). As an example of the convergence of $E^{(2)}$
with the number of partial waves $l$, consider the ground $4s$ state. Calculations of $E^{(2)}$ with $l_{\text{max}}$ =
6 and  8 yield
 $E^{(2)}(4s)$ = $-4726.2$ and
$-4743.7$~cm$^{-1}$, respectively. Extrapolation of these calculations yields $-4786.3$ and $-4786.9$~cm$^{-1}$, respectively. Therefore, the
numerical uncertainty in the  second-order value $E^{(2)}(4s)$ is 0.6~cm$^{-1}$. It should be noted that the 17.5~cm$^{-1}$ contribution from partial
waves with $l>6$ for the $4s$ state is the largest among all states considered in Table~\ref{tab1};  smaller (about $4 - 6$~cm$^{-1}$) contributions
are obtained for the $3d$, $4p$, and $4d$ states and  much smaller contributions ($0.5-1.5$~cm$^{-1}$) are obtained for the $n$ = 6 states.

Owing to complexity of the all-order calculations, we restrict $l\leq l_{\text{max}} = 6$ in the $E^\text{SD}$ calculation. The second-order
contribution dominates $E^\text{SD}$; therefore, we can use the extrapolated value of the $E^{(2)}$ described above to account for the contributions
of the higher partial waves. The partial waves $l \leq 6$ are also used in the calculation of $E^{(3)}$.
  Since the asymptotic
$l$-dependence of the second- and third-order energies are similar (both fall off as $l^{-4}$), we use the second-order remainder as a guide to
estimate the remainder in the third-order contribution. The term $E^{(3)}_\text{{extra}}$ in Table~\ref{tab1}, which accounts for the part of the
third-order MBPT energy missing from the SD expression for the energy,  is smaller than $E^{(3)}$  by an order of magnitude for the states considered
here.

The column labeled $\delta E^\text{SD}$ in Table~\ref{tab1} gives differences between our {\it ab initio}
 results and the experimental values \cite{nist-web}.
The SD results agree significantly better with measured values than do the third-order MBPT results (the ratio of
$\delta E^{(3)}$/$\delta E^\text{SD}$ is about 10 for some of cases), illustrating the importance of fourth and
higher-order correlation corrections.

We provide recommended values for the $7p_{1/2}$
 and $7p_{3/2}$ energies in Table~\ref{tab1} in the separate rows. We estimate these values to be accurate to about 3~cm$^{-1}$.
 Our recommended values for the
$8p_{1/2}$
 and $8p_{3/2}$ energies are $-10221$~cm$^{-1}$
and $-10209$~cm$^{-1}$, respectively.

\begin{table*}
\caption{\label{tab-dip} Recommended values of the reduced electric-dipole matrix elements in atomic units.  The
 first-order, second-order,  third-order MBPT, and all-order SD and SDpT  values are listed; the label ``sc'' indicates the scaled values. Final recommended values and their
uncertainties are given in the $Z^\text{{final}}$ column. The last column gives relative uncertainties of the final values in \%. Absolute values are
given.}
\begin{ruledtabular}
\begin{tabular}{llrrrrrrrrc}
\multicolumn{2}{c}{Transition}& \multicolumn{1}{c}{$Z^{{\rm DF}}$ }& \multicolumn{1}{c}{$Z^{({\rm DF}+2)}$ }& \multicolumn{1}{c}{$Z^{({\rm DF}+2+3)}$
}& \multicolumn{1}{c}{$Z^\text{{SD}}$ }& \multicolumn{1}{c}{$Z_{\rm sc}^\text{{(SD)}}$ }& \multicolumn{1}{c}{$Z^\text{{SDpT}}$ }&
\multicolumn{1}{c}{$Z_{\rm sc}^\text{{SDpT}}$ }& \multicolumn{1}{c}{$Z^\text{{final}}$ }&
\multicolumn{1}{c}{Unc. (\%)}\\
\hline
  $ 4s_{1/2}$&$   4p_{1/2}$&    3.2012&    3.0045&    2.8826&     2.8978&    2.9071&     2.9131&     2.9071&    2.898(13) &    0.45\\
  $ 4s_{1/2}$&$   4p_{3/2}$&    4.5269&    4.2499&    4.0773&     4.0989&    4.1119&     4.1204&     4.1119&    4.099(18) &    0.45\\
  $ 5s_{1/2}$&$   4p_{1/2}$&    2.1084&    2.1458&    2.0799&     2.0660&    2.0735&     2.0706&     2.0717&    2.073(11) &    0.51\\
  $ 5s_{1/2}$&$   4p_{3/2}$&    3.0142&    3.0653&    2.9752&     2.9551&    2.9647&     2.9614&     2.9622&    2.965(14) &    0.46\\
  $ 5s_{1/2}$&$   5p_{1/2}$&    6.4426&    6.3777&    6.1965&     6.2195&    6.2297&     6.2392&     6.2287&    6.23(1) &    0.16\\
  $ 5s_{1/2}$&$   5p_{3/2}$&    9.1006&    9.0095&    8.7523&     8.7850&    8.7999&     8.8130&     8.7984&    8.80(1) &    0.17\\
  $ 6s_{1/2}$&$   4p_{1/2}$&    0.5798&    0.6002&    0.5837&     0.5817&    0.5820&     0.5825&     0.5827&   0.582(1) &    0.18\\
  $ 6s_{1/2}$&$   4p_{3/2}$&    0.8239&    0.8522&    0.8291&     0.8264&    0.8267&     0.8276&     0.8278&   0.826(2) &    0.18\\
  $ 6s_{1/2}$&$   5p_{1/2}$&    4.4346&    4.4565&    4.3672&     4.3504&    4.3606&     4.3529&     4.3529&   4.361(8) &    0.18\\
  $ 6s_{1/2}$&$   5p_{3/2}$&    6.3311&    6.3601&    6.2388&     6.2144&    6.2267&     6.2177&     6.2159&    6.23(1) &    0.17\\
  $ 6s_{1/2}$&$   6p_{1/2}$&   10.7169&   10.6885&   10.4530&    10.4853&   10.4977&    10.5108&    10.4952&   10.50(1) &    0.12\\
  $ 6s_{1/2}$&$   6p_{3/2}$&   15.1300&   15.0906&   14.7554&    14.8016&   14.8201&    14.8379&    14.8166&   14.82(2) &    0.12\\
  $ 7s_{1/2}$&$   4p_{1/2}$&    0.3140&    0.3276&    0.3193&     0.3183&    0.3182&     0.3186&     0.3187&  0.3183(6) &    0.20\\
  $ 7s_{1/2}$&$   4p_{3/2}$&    0.4456&    0.4644&    0.4530&     0.4515&    0.4514&     0.4520&     0.4520&  0.4515(9) &    0.20\\
  $ 7s_{1/2}$&$   5p_{1/2}$&    1.1176&    1.1301&    1.1057&     1.1056&    1.1061&     1.1075&     1.1073&   1.106(1) &    0.12\\
  $ 7s_{1/2}$&$   5p_{3/2}$&    1.5845&    1.6020&    1.5676&     1.5675&    1.5681&     1.5702&     1.5699&   1.568(2) &    0.13\\
  $ 7s_{1/2}$&$   6p_{1/2}$&    7.4782&    7.4912&    7.3825&     7.3492&    7.3629&     7.3587&     7.3582&   7.363(5) &    0.06\\
  $ 7s_{1/2}$&$   6p_{3/2}$&   10.6693&   10.6877&   10.5391&    10.4913&   10.5071&    10.5040&    10.5009&  10.507(6) &    0.06\\
  $ 7s_{1/2}$&$   7p_{1/2}$&    16.0333&  16.0184&   15.7262&     15.7723&  15.7894 &   15.8043 &   15.7856 &  15.79(1) &    0.09\\
  $ 7s_{1/2}$&$   7p_{3/2}$&    22.6281&  22.6084&   22.1909&     22.2570&  22.2812 &   22.3027 &    22.2758 & 22.28(2) &    0.10\\
 $  8s_{1/2}$&$   5p_{1/2}$&    0.5862&    0.5941&    0.5827&     0.5829&    0.5830 &    0.5840 &    0.5839 &   0.583(1) &    0.17\\
 $  8s_{1/2}$&$   5p_{3/2}$&    0.8299&    0.8411&    0.8248&     0.8252&    0.8253 &    0.8267 &    0.8265 &   0.825(1) &    0.17\\
 $  8s_{1/2}$&$   6p_{1/2}$&    1.7969&    1.8017&    1.7732&     1.7738&    1.7743 &    1.7770 &    1.7767 &   1.774(3) &    0.15\\
 $  8s_{1/2}$&$   6p_{3/2}$&    2.5448&    2.5547&    2.5112&     2.5123&    2.5129 &    2.5167 &    2.5163 &   2.513(4) &    0.15\\
  $ 8s_{1/2}$&$   7p_{1/2}$&    11.2499&  11.2611&    11.1339&   11.0842 &  11.0946 &   11.0951 &   11.0879 &  11.095(7) &    0.06\\
  $ 8s_{1/2}$&$   7p_{3/2}$&    16.0443&  16.0563&    15.8877&   15.8166 &  15.8313 &   15.8309 &   15.8219 &  15.831(9) &    0.06\\
  $ 8s_{1/2}$&$   8p_{1/2}$&    22.3941&  22.3854&    22.0348&   22.0969 &         &     22.1359 &          &   22.14(4) &    0.18\\
  $ 8s_{1/2}$&$   8p_{3/2}$&    31.5981&  31.5878&    31.0854&   31.1745 &         &     31.2302 &          &   31.23(6) &    0.18\\
$ 3d_{3/2} $&$  4p_{1/2}$&      3.0825 &   2.9296&     2.2998&   2.4173 &     2.4636 &    2.4677 &   2.4503 &    2.464(13) &  0.54\\
$ 3d_{3/2} $&$  4p_{3/2}$&      1.3764 &   1.3088&     1.0260&   1.0788 &     1.0996 &    1.1014 &   1.0937 &   1.100(6) &  0.54\\
$ 3d_{3/2} $&$  4f_{5/2}$&      2.6059 &   2.5228&     1.6763&    1.8660 &    1.9265 &    1.9265 &   1.9051 &    1.927(21) &  1.11\\
$ 3d_{3/2} $&$  5f_{5/2}$&      1.5216 &   1.4530&     1.1073&    1.1655 &    1.1917 &    1.1937 &   1.1846 &   1.192(7) &  0.59\\
$ 3d_{5/2} $&$  4p_{3/2}$&      4.1348 &   3.9311&     3.0882&    3.2452 &    3.3063 &    3.3127 &   3.2884 &    3.306(18) &  0.54\\
$ 3d_{5/2} $&$  4f_{5/2}$&      0.6976 &   0.6751&     0.4502&    0.5005 &    0.5163 &    0.5166 &   0.5106 &   0.516(6) &  1.10\\
$ 3d_{5/2} $&$  4f_{7/2}$&      3.1201 &   3.0192&     2.0134&    2.2382 &    2.3090 &    2.3102 &   2.2835 &    2.309(25) &  1.10\\
$ 3d_{5/2} $&$  5f_{5/2}$&      0.4072 &   0.3888&     0.2970&    0.3124 &    0.3192 &    0.3199 &   0.3174 &   0.319(2) &  0.59\\
$ 3d_{5/2} $&$  5f_{7/2}$&      1.8214 &   1.7388&     1.3284&    1.3972 &    1.4278 &    1.4308 &   1.4193 &   1.428(8) &  0.59\\
 $  4d_{3/2}$&$   4p_{1/2}$&    4.2159 &   4.1495&     4.4121&    4.2636 &    4.2818 &    4.2565 &      4.2821 &    4.28(3) &  0.59\\
 $  4d_{3/2}$&$   4p_{3/2}$&    1.8990 &   1.8686&     1.9867&    1.9203 &    1.9280 &    1.9171 &     1.9281 &    1.93(1) &   0.57\\
 $  4d_{3/2}$&$   4f_{5/2}$&   11.9755&   11.8972&    11.3300&    11.3461 &  11.3552 &   11.4033 &   11.3479 &   11.36(5) &    0.42\\
 $  4d_{3/2}$&$   5p_{1/2}$&    8.0685&    8.0369&     7.2981&     7.4342 &   7.4325 &    7.4916 &    7.4704 &    7.43(6) &    0.79\\
 $  4d_{3/2}$&$   5p_{3/2}$&    3.6014&    3.5878&     3.2554&     3.3168 &   3.3162 &    3.3426 &    3.3320 &    3.32(3) &    0.79\\
 $  4d_{5/2}$&$   4p_{3/2}$&    5.6913&    5.6015&     5.9527&     5.7547 &    5.7786 &    5.7451 &    5.7791 &    5.78(3) &    0.58\\
 $  4d_{5/2}$&$   5p_{3/2}$&   10.8150&   10.7724&     9.7806&     9.9632 &   9.9600 &   10.0403 &   10.0091 &    9.96(8) &    0.81\\
 $  4d_{5/2}$&$   4f_{5/2}$&    3.2020&    3.1809&     3.0300&     3.0342 &   3.0364 &    3.0494 &    3.0345 &    3.04(1) &    0.43\\
 $  4d_{5/2}$&$   4f_{7/2}$&   14.3198&   14.2254&    13.5508&    13.5696 &  13.5793 &   13.6376 &   13.5707 &   13.58(6) &    0.43\\
 $  5d_{3/2}$&$   5p_{1/2}$&    7.1941&    7.1786&     7.6248&    7.4598 &    7.4916 &    7.4316 &    7.4808 &    7.49(6) &    0.80\\
 $  5d_{3/2}$&$   5p_{3/2} $&   3.2436&    3.2357&     3.4371&    3.3630 &    3.3763 &    3.3503 &    3.3715 &    3.38(3) &    0.77\\
 $  5d_{3/2}$&$   4f_{5/2}$&    7.9026&    7.9162&     8.7463&     8.5477 &   8.5564 &    8.4871 &    8.5512 &    8.56(7) &    0.81\\
 $  5d_{5/2}$&$   5p_{3/2}$&    9.7176&    9.6962&    10.2951&    10.0746 &  10.1162 &  10.0367 &     10.1018 &   10.12(8) &    0.79\\
 $  5d_{5/2}$&$   4f_{5/2}$&    2.1094&    2.1135&    2.3342&     2.2815 &   2.2841 &    2.2653 &    2.2828 &    2.28(2) &    0.82\\
 $  5d_{5/2}$&$   4f_{7/2}$&    9.4333&    9.4515&   10.4388&    10.2029 &  10.2149 &   10.1307 &   10.2088 &   10.21(8) &    0.82\\
 $  5d_{3/2}$&$   5f_{5/2}$&   20.4001&   20.3664&   19.0555&    19.1963 &  19.2049 &   19.3158 &   19.1938 &    19.2(1) &    0.58\\
 $  5d_{5/2}$&$   5f_{5/2}$&    5.4558&    5.4462&    5.0977&     5.1350 &   5.1367 &    5.1668 &    5.1337 &    5.14(3) &    0.59\\
 $  5d_{5/2}$&$   5f_{7/2}$&   24.3991&   24.3564&   22.7979&    22.9645 &  22.9721 &   23.1067 &   22.9588 &    23.0(1) &    0.59\\
 $  6d_{3/2}$&$   5p_{1/2}$&    2.3088&    2.2923&    2.2439&     2.2293 &   2.2284 &    2.2369 &    2.2368 &   2.228(9) &    0.39\\
 $  6d_{3/2}$&$   5p_{3/2}$&    1.0338&    1.0263&    1.0032&     0.9969 &   0.9965 &    1.0005 &    1.0004 &   0.996(4) &    0.40\\
 $  6d_{5/2}$&$   5p_{3/2}$&    3.1018&    3.0793&    3.0112&     2.9922 &   2.9908 &    3.0027 &    3.0025 &    2.99(1) &    0.40\\
 \end{tabular}
\end{ruledtabular}
\end{table*}

 %*************************************
\section{Electric-dipole matrix elements, oscillator strengths, transition
rates, and lifetimes in Ca~II}

\subsection{Electric-dipole matrix elements}
\label{E1}

In Table \ref{tab-dip}, we list our recommended values for 58 E1 $ns-n^{\prime}p$ and $nd-n^{\prime}p$ transitions. We note that we have calculated
over 500 E1 matrix elements to evaluate polarizabilities and BBR shift presented in this work. We refer to these values as ``best set'' of the matrix
elements. We list only the matrix elements that give significant contributions to the atomic properties calculated in the other sections. To evaluate
the uncertainties of these values, we carried out several calculations in different approximations. To demonstrate the size of the second, third, and
higher-order correlation corrections, we list the lowest-order Dirac-Fock (DF) $Z^{{\rm DF}}$, second-order $Z^{({\rm DF}+2)}$, and third-order
$Z^{({\rm DF}+2+3)}$ values in the first three numerical columns of Table~\ref{tab-dip}. The absolute values in atomic units ($a_0e$) are given in
all cases. The many-body perturbation theory (MBPT) calculations are carried out following the method described in Ref.~\cite{adndt-96}.
 The values  $Z^{({\rm DF}+2)}$ are obtained as the sum of
 the second-order correlation correction $Z^{(2)}$
   and  the DF matrix elements $Z^{\rm DF}$. The second-order Breit corrections $B^{(2)}$ are very
    small in comparison with  the second-order
Coulomb corrections $Z^{(2)}$ (the ratio of $B^{(2)}$ to $Z^{(2)}$ are about 1\%--2\%).
 The third-order matrix elements $Z^{({\rm DF}+2+3)}$ include the DF values,
 the second-order $Z^{(2)}$ results,
and the third-order $Z^{(3)}$  correlation correction. $Z^{(3)}$ includes random-phase-approximation  terms (RPA) iterated  to all orders, Brueckner
orbital (BO) corrections, the structural radiation,  and normalization terms (see \cite{adndt-96} for definition of these terms).

Next four columns contain four different all-order calculations. \textit{Ab initio} electric-dipole matrix elements evaluated in the all-order  SD
(single-double)  and SDpT approximations (single-double all-order method including partial triple excitations \cite{mar-pol-99}) are given  in
columns labeled $Z^\text{SD}$ and $Z^\text{SDpT}$
 of Table~\ref{tab-dip}. The SD and SDpT matrix elements
$Z^\text{SD}$ include $Z^{(3)}$ completely, along with important fourth- and higher-order corrections. The fourth-order corrections omitted from the
SD matrix elements were discussed  by \citet{der-4}.  Difference between the $Z^\text{SD}$ and $Z^\text{SDpT}$ values is about 0.5~\% - 2.0~\%.

\begin{table}
\caption{\label{comp1} Comparison of the present values of E1 dipole matrix elements with RCC calculations of Ref.~\cite{das-09}. The uncertainties
in our values represent our best estimate of all possible sources of uncertainties, i.e. they give estimated boundary values of these recommended
results. The uncertainties in Ref.~\cite{das-09} values are numerical uncertainties resulting from the  use of incomplete basis sets. Absolute values
in atomic units are given.}
\begin{ruledtabular}
\begin{tabular}{lrrrr}
\multicolumn{1}{c}{Transition}& \multicolumn{1}{c}{Present}& \multicolumn{3}{c}{Ref.~\cite{das-09}}\\
\multicolumn{1}{c}{}& \multicolumn{1}{c}{}& \multicolumn{1}{c}{STOs}&
\multicolumn{1}{c}{GTOs}&\multicolumn{1}{c}{Final}
\\
\hline
 $4p_{1/2}-4s      $&    2.898(13) &  2.86 & 2.90 & 2.88(1)\\
 $4p_{3/2}-4s      $&    4.099(18) &  4.02 & 4.09 & 4.03(1) \\
 $4p_{1/2}-3d_{3/2}$&    2.464(13) &  2.50 & 2.41 & 2.40(2)\\
 $4p_{3/2}-3d_{3/2}$&    1.100(6)  &  1.12 & 1.09 & 1.09(1) \\
 $4p_{3/2}-3d_{5/2}$&    3.306(18) &  3.36 & 3.28 & 3.22(4) \\
\end{tabular}
\end{ruledtabular}
\end{table}

\begin{table*}
\caption{\label{comp2} Comparison of the ratios $R=d^2_1/d^2_2$ of the squares of the E1 matrix elements calculated in
the present work in several approximations with results of  Ref.~\cite{das-09}.}
\begin{ruledtabular}
\begin{tabular}{cccccccccccccc}
       $d_1$      &        $d_2$            &\multicolumn{7}{c}{Present work}  &&\multicolumn{3}{c}{Ref.~\cite{das-09}}\\
  &&DF   &    MBPT2 &MBPT3 &    SD  &  SDsc &     SDpT &   SDpTsc &&    STOs&GTOs&Final\\
       \hline
$4s- 4p_{3/2}$&        $4s- 4p_{1/2}$     &  2.000 &  2.001&   2.001 &  2.001 &   2.001 &   2.001 &   2.001&& 1.976
&1.989  &  1.958(17)
\\
$4p_{1/2}-3d_{3/2}$&   $4p_{3/2}-3d_{3/2}$&  5.02  &  5.01 &   5.02  &  5.02  &   5.02  &   5.02  &   5.02 &&   4.98 &   4.89 &  4.85(12)\\
$4p_{3/2}-3d_{5/2}$&   $4p_{3/2}-3d_{3/2}$&  9.02  &  9.02 &   9.06  &  9.05  &   9.04  &   9.05  &   9.04 &&   9.00 &
9.06 &8.73(27)
   \\
\end{tabular}
\end{ruledtabular}
\end{table*}

We have developed some general criteria to establish the final values for all transitions and evaluate uncertainties owing to the need to analyze a
very large number of transitions. To evaluate the uncertainties of our  matrix element values and to provide recommended values, we carried out
semi-empirical evaluation of the missing correlation corrections using the scaling procedure described below.

The matrix elements of  any one-body operator $Z = \sum_{ij} z_{ij}\ a^\dagger_i a_j$ are obtained within the framework of the SD all-order method as
\begin{equation}
Z_{wv}=\frac{\langle \Psi_w |Z| \Psi_v \rangle}{\sqrt{\langle \Psi_v | \Psi_v \rangle \langle \Psi_w | \Psi_w \rangle}},  \label{matrix}
\end{equation}
where $|\Psi_v\rangle$ and $|\Psi_w\rangle$ are given by the expansion
\begin{eqnarray}
&& |\Psi_v \rangle = \left[ 1 + \sum_{ma} \, \rho_{ma} a^\dagger_m a_a + \frac{1}{2} \sum_{mnab} \rho_{mnab}
a^\dagger_m a^\dagger_n a_b a_a
\right.  \notag \\
&& + \left. \sum_{m \neq v} \rho_{mv} a^\dagger_m a_v + \sum_{mna} \rho_{mnva} a^\dagger_m a^\dagger_n a_a a_v \right]| \Psi_v^{(0)}\rangle ,
\label{eqr}
\end{eqnarray}
and $|\Psi_v^{(0)}\rangle$ is the lowest-order atomic state vector. In Eq.~(\ref{eqr}), the indices $m$ and $n$ range over all possible virtual
states while indices $a$ and $b$ range over all occupied core states. The quantities $\rho_{ma}$, $\rho_{mv}$ are
single-excitation coefficients for core and valence electrons and $%
\rho_{mnab}$ and $\rho_{mnva}$ are double-excitation coefficients for core and valence electrons, respectively.  In the SD approximation, the
resulting expression for the numerator of Eq.~(\ref{matrix}) consists of the sum of the DF matrix element $z_{wv}$ and 20 other terms that are linear
or quadratic functions of the excitation coefficients. The all-order method yielded results for the properties of alkali-metal atoms and many other
monovalent systems~\cite{mar-pol-99,mg,sr,ba,review07} in excellent agreement with experiment. For example, the SD results for the primary
$ns-np_{j}$ E1 matrix elements of alkali-metal atoms agree with experiment to 0.1\%-0.5\% \cite{mar-pol-99}. However, triple corrections are
important for many of the $nd-n^{\prime}p$ matrix elements and have to be included. Our \textit{ab initio} SDpT values include corrections to the
equations for the valence excitation coefficients $\rho_{mv}$ and valence energy. These corrections arise from the addition of the valence triple
excitations to the wave function given by Eq.~(\ref{eqr}).

We find that only two terms give dominant contributions for all matrix elements considered in this work:
\begin{equation}
Z^{(a)} =  \sum\limits_{ma} \left( z_{am} \tilde{\rho}_{wmva} + z_{ma} \tilde{\rho}^{*}_{vmwa} \right)
\end{equation}
or
\begin{equation}
Z^{(c)} =  \sum\limits_{m} \left( z_{wm} \rho_{mv} + z_{mv} \rho^{*}_{mw} \right),
\end{equation}
where $\tilde{\rho}_{mnab}=\rho_{mnab}-\rho_{nmab}$ and $z_{wv}$ are lowest-order matrix elements of the electric-dipole operator. For most of the
transitions considered in this work, term $Z^{(c)}$  is the dominant term. In many cases, it is overwhelmingly dominant (by a factor of 3 or more).
To evaluate missing corrections to this term, we need to improve the values of the valence single-excitation coefficients $\rho_{mv}$ \cite{quad}.
 These excitation coefficients are closely related to the correlation energy $\delta E_v$. If we
introduce the self-energy operator $\Sigma_{mv}$ (also referred to as correlation potential in some works)  as
\begin{equation}
\Sigma_{mv}=\left( \widetilde{\epsilon}_v - \epsilon_m\right) \rho_{mv},
\end{equation}
then the correlation energy would correspond to the diagonal term $\Sigma_{vv}$ \cite{CI}. Therefore, the omitted correlation correction can be
estimated by adjusting the single-excitation coefficients $\rho_{mv}$ to the experimentally known value of the valence correlation energy, and then
re-calculating the matrix elements using Eq.~(\ref{matrix}) with the modified coefficients \cite{quad}
\begin{equation}
\rho_{mv}^{\prime}=\rho_{mv} \frac{\delta E_v^{\textrm{expt}}}{\delta E_v^{\textrm{theory}}}. \label{scale}
\end{equation}
 The $\delta E_v^{\textrm{expt}}$ is defined as the experimental
 energy \cite{nist-web}
minus the lowest order DF energy $\epsilon_v$.  We note that it is a rather complicated procedure that involves complete recalculation of the matrix
elements with new values of the valence excitation coefficients. The scaling factors depend on the correlation energy given by the particular
calculation. Therefore, the scaling factors are different for the SD and SDpT calculations, and these values have to be scaled separately. Generally,
scaled SD and SDpT values are close together, as expected. The corresponding results are listed in Table~\ref{tab-dip} with
 subscript ``sc".

 The term $Z^{(a)}$
is not  corrected by the scaling procedure. However, it is dominant for very few transitions that give significant contributions to the atomic
properties considered in this work. Essentially, the only large matrix elements where term $Z^{(c)}$ is not dominant are $4s-4p$ and $5s-5p$. In both
of these cases, term  $Z^{(c)}$ is still of the same order magnitude as the term $Z^{(a)}$. Therefore, we can establish the recommended set of values
and their uncertainties based on the ratio $R = Z^{(c)}/Z^{(a)}$. We take the final value to be SD scaled if $R>1$. Otherwise, we use SD as the final
value. If $0.5< R <1.5$, we evaluate the uncertainty in term $Z^{(c)}$ as the maximum difference of the final value and the other three all-order
values from the  SD, SDpT, SDsc, and SDpTsc set. Then, we assume that the uncertainty of all the other terms does not exceed this value and add two
uncertainties in quadrature.
 If $1.5< R <3$, we evaluate the final uncertainty as the max(SDsc-SD, SDsc-SDpT, SDsc-SDpTsc). If the term $Z^{(c)}$ strongly dominates
 and $R>3$, we evaluate the final uncertainty as max(SDsc-SDpT, SDsc-SDpTsc).
 We note that we have conducted numerous comparisons of all available data on various  properties of many different monovalent
systems with different types of experiments in many other works (see
\cite{mar-bbr-07,mar-pol-99,review07,mg,sr,ba,quad,safr-sr2,safr-ba-10,ca-3d-expt-05,tl-05,n1,n2,n3,n4,n5} and references therein)  and found that
such procedures do not underestimate the uncertainties. If fact, they may somewhat overestimate the uncertainties in some cases.

 The last column of Table~\ref{tab-dip} gives relative uncertainties
of the final values $Z^{\rm final}$ in \%.
  We find that
the uncertainties are 0.2-0.5\% for most of the transitions. Larger uncertainties (0.8\%) occur for some of the
transitions such as $5d_{j}-4f_{j'}$. Our final results and their uncertainties
 are used to calculate the recommended values of the transition rates, oscillator strengths,  lifetimes, and polarizabilities as well as
 evaluate the uncertainties of these results.

Two most recent calculations of the E1 matrix elements between the low-lying states were carried out by Sahoo et al. \cite{das-09} using the
relativistic coupled-cluster method (RCC) and by Mitroy and Zhang \cite{mitroy-08} using non-relativistic configuration interaction with a
semi-empirical core potential (CICP)  approach. Ref.~\cite{das-09} includes comparison with earlier MBPT calculations   \cite{nd-wrj-91,ca-liaw-95}.
Since \cite{nd-wrj-91,ca-liaw-95} only include low-order MBPT
 corrections, these calculations are substantially less complete than all-order coupled-cluster method used in our work and
   Ref.~\cite{das-09}. Therefore, we focus our discussion on the comparison of the present results with those of
   \cite{das-09}. Since  Ref.~\cite{mitroy-08} presents non-relativistic calculations and lists oscillator strengths rather than matrix elements,
   we  compare their results  with our $j$-averaged oscillator strengths in the next section.
We note that we use the same method as \cite{mar-bbr-07} and our results for the transitions listed in
\cite{mar-bbr-07} are the same. Therefore, we do not include separate comparison with that work.

\begin{table*}
\caption{\label{tab-osc1} Wavelengths $\lambda$ (\AA), transition rates $A$ (s$^{-1}$) and oscillator strengths ($f$) for transitions in Ca~II
calculated using our recommended values of reduced electric-dipole matrix elements $Z^\text{{final}}$ and their uncertainties. The relative
uncertainties in the values of transition rates and oscillator strengths are the same. They are listed in column ``Unc.'' in \%. Numbers in brackets
represent powers of 10.}
\begin{ruledtabular}
\begin{tabular}{llrlllllrlll}
\multicolumn{2}{c}{Transition} & \multicolumn{1}{c}{$\lambda$} & \multicolumn{1}{c}{$A$} & \multicolumn{1}{c}{$f$} & \multicolumn{1}{c}{Unc.} &
 \multicolumn{2}{c}{Transition} &
\multicolumn{1}{c}{$\lambda$} & \multicolumn{1}{c}{$A$} & \multicolumn{1}{c}{$f$} &
\multicolumn{1}{c}{Unc.} \\
\hline
 $4s_{1/2}$&$   4p_{1/2} $&   3969.6&   1.360[8]&    3.213[-1]& 0.90 & $4d_{3/2}$&$   4f_{5/2} $&    8914.5&   6.146[7]&    1.098[ 0]& 0.84\\
 $4s_{1/2}$&$   4p_{3/2} $&   3934.8&   1.397[8]&    6.485[-1]& 0.90 & $4d_{5/2}$&$   4f_{5/2} $&    8929.8&   4.372[6]&    5.227[-2]& 0.86\\
 $5s_{1/2}$&$   5p_{1/2} $&  11953.0&   2.302[7]&    4.931[-1]& 0.32 & $4d_{5/2}$&$   4f_{7/2} $&    8929.8&   6.559[7]&    1.045[ 0]& 0.86\\
 $5s_{1/2}$&$   5p_{3/2} $&  11842.2&   2.362[7]&    9.932[-1]& 0.34 & $5d_{3/2}$&$   5f_{5/2} $&   18824.7&   1.867[7]&    1.488[ 0]& 1.16\\
 $6s_{1/2}$&$   6p_{1/2} $&  26265.3&   6.161[6]&    6.372[-1]& 0.24 & $5d_{5/2}$&$   5f_{5/2} $&   18855.6&   1.329[6]&    7.084[-2]& 1.18\\
 $6s_{1/2}$&$   6p_{3/2} $&  26013.7&   6.320[6]&    1.282[ 0]& 0.24 & $5d_{5/2}$&$   5f_{7/2} $&   18855.6&   1.994[7]&    1.417[ 0]& 1.18\\[0.4pc]
 $4p_{1/2}$&$   5s_{1/2} $&   3707.1&   8.550[7]&    1.761[-1]& 1.02 & $4p_{1/2}$&$   4d_{3/2} $&    3159.8&   2.944[8]&    8.812[-1]& 1.18\\
 $4p_{3/2}$&$   5s_{1/2} $&   3738.0&   1.705[8]&    1.786[-1]& 0.92 & $4p_{3/2}$&$   4d_{3/2} $&    3182.2&   5.843[7]&    8.871[-2]& 1.14\\
 $4p_{1/2}$&$   6s_{1/2} $&   2198.5&   3.226[7]&    2.338[-2]& 0.36 & $4p_{3/2}$&$   4d_{5/2} $&    3180.3&   3.506[8]&    7.973[-1]& 1.16\\
 $4p_{3/2}$&$   6s_{1/2} $&   2209.3&   6.416[7]&    2.347[-2]& 0.36 & $4p_{1/2}$&$   5d_{3/2} $&    2103.9&   8.102[7]&    1.075[-1]& 2.58\\
 $4p_{1/2}$&$   7s_{1/2} $&   1843.1&   1.639[7]&    8.349[-3]& 0.40 & $4p_{3/2}$&$   5d_{3/2} $&    2113.8&   1.592[7]&    1.067[-2]& 2.62\\
 $4p_{3/2}$&$   7s_{1/2} $&   1850.7&   3.258[7]&    8.365[-3]& 0.40 & $4p_{3/2}$&$   5d_{5/2} $&    2113.4&   9.580[7]&    9.623[-2]& 2.62\\
 $4p_{1/2}$&$   8s_{1/2} $&   1691.8&   9.528[6]&    4.088[-3]& 0.42 & $5p_{1/2}$&$   5d_{3/2} $&    8204.0&   5.149[7]&    1.039[ 0]& 1.60\\
 $4p_{3/2}$&$   8s_{1/2} $&   1698.2&   1.893[7]&    4.092[-3]& 0.42 & $5p_{3/2}$&$   5d_{3/2} $&    8257.0&   1.026[7]&    1.048[-1]& 1.54\\
 $5p_{1/2}$&$   6s_{1/2} $&   9857.5&   2.011[7]&    2.930[-1]& 0.36 & $5p_{3/2}$&$   5d_{5/2} $&    8251.1&   6.152[7]&    9.419[-1]& 1.58\\
 $5p_{3/2}$&$   6s_{1/2} $&   9934.1&   4.007[7]&    2.964[-1]& 0.34 & $5p_{1/2}$&$   6d_{3/2} $&    5002.9&   2.009[7]&    1.508[-1]& 0.78\\
 $5p_{1/2}$&$   7s_{1/2} $&   5286.7&   8.388[6]&    3.515[-2]& 0.24 & $5p_{3/2}$&$   6d_{3/2} $&    5022.5&   3.970[6]&    1.501[-2]& 0.80\\
 $5p_{3/2}$&$   7s_{1/2} $&   5308.7&   1.665[7]&    3.517[-2]& 0.26 & $5p_{3/2}$&$   6d_{5/2} $&    5021.4&   2.386[7]&    1.353[-1]& 0.80\\
 $5p_{1/2}$&$   8s_{1/2} $&   4207.4&   4.623[6]&    1.227[-2]& 0.34 &                          &          &           &             &    \\
 $5p_{3/2}$&$   8s_{1/2} $&   4221.3&   9.173[6]&    1.225[-2]& 0.34 & $3d_{3/2}$&$   4p_{1/2} $&    8664.5&   9.452[6]&    5.319[-2]& 1.08\\
 $5p_{1/2}$&$   9s_{1/2} $&   3740.4&   2.859[6]&    5.997[-3]& 0.38 & $3d_{3/2}$&$   4p_{3/2} $&    8500.4&   9.972[5]&    1.080[-2]& 1.08\\
 $5p_{3/2}$&$   9s_{1/2} $&   3751.4&   5.670[6]&    5.982[-3]& 0.38 & $3d_{5/2}$&$   4p_{3/2} $&    8544.4&   8.876[6]&    6.477[-2]& 1.08\\
 $5p_{1/2}$&$  10s_{1/2} $&   3486.6&   1.899[6]&    3.462[-3]& 0.54 & $4d_{3/2}$&$   5p_{1/2} $&   27072.6&   2.820[6]&    1.550[-1]& 1.58\\
 $5p_{3/2}$&$  10s_{1/2} $&   3496.2&   3.768[6]&    3.453[-3]& 0.54 & $4d_{3/2}$&$   5p_{3/2} $&   26510.9&   2.990[5]&    3.150[-2]& 1.58\\
 $6p_{1/2}$&$   7s_{1/2} $&  20147.6&   6.715[6]&    4.087[-1]& 0.12 & $4d_{5/2}$&$   5p_{3/2} $&   26646.6&   2.656[6]&    1.885[-1]& 1.60\\
 $6p_{3/2}$&$   7s_{1/2} $&  20298.3&   1.337[7]&    4.130[-1]& 0.12 &                          &          &           &             &    \\
 $6p_{1/2}$&$   8s_{1/2} $&  10187.5&   3.016[6]&    4.693[-2]& 0.30 & $4f_{5/2}$&$   5d_{3/2} $&   21434.8&   3.766[6]&    1.729[-1]& 1.62\\
 $6p_{3/2}$&$   8s_{1/2} $&  10225.8&   5.983[6]&    4.689[-2]& 0.30 & $4f_{5/2}$&$   5d_{5/2} $&   21394.9&   1.799[5]&    1.235[-2]& 1.64\\
           &              &         &           &             &      & $4f_{7/2}$&$   5d_{5/2} $&   21394.9&   3.598[6]&    1.852[-1]& 1.64\\
\end{tabular}
\end{ruledtabular}
\end{table*}

Our final values are compared with RCC calculations of Sahoo et al. \cite{das-09} in Table~\ref{comp1}. We discuss this comparison in significant
detail since both calculations are carried out using the couple-cluster method but differ significantly in its implementation.  All E1 transitions
listed in \cite{das-09} are included. The results of \cite{das-09} listed in columns labeled ``STOs'' and ''GTOs'' are obtained by two different
calculations, one with Slater-type orbitals and another with Gaussian-type orbitals, respectively. In both cases,  the number of partial waves was
restricted to $l_{max}=4$, i.e. only   $s, p, d, f$, and $g$ orbitals were included. Table~\ref{comp1} illustrates significant basis set dependence
in the results of Sahoo et al. \cite{das-09}, 1.4-1.7\% for the $4s-4p$ transitions and 2.4-3.8\% for the $3d-4p$ transitions. Moreover, different
fine-stricture components have different basis set dependencies. It is not clear how that may be possible unless some additional  basis set
optimization was carried out in different way for all transitions. The final results from \cite{das-09} include corrections from higher symmetry
orbitals carried out using MBPT(2); the changes between the GTOs/STOs values and final recommended results range from 0.7\% to 4.3\%. Unfortunately,
the Ref.~\cite{das-09} does not explicitly state what terms are accounted for by MBPT(2). In the standard formulation of the perturbation theory
\cite{adndt-96}, second-order perturbation theory contains only random-phase approximation terms which are relatively small for the $3d-4p$
transitions. The main contribution of the higher partial waves only appears starting from the third order and comes from so-called Brueckner-orbital
terms \cite{adndt-96}. Therefore, using  the MBPT(2) to evaluate higher symmetry contributions should severely underestimate these terms.  The
uncertainties of the final values from \cite{das-09} are numerical uncertainties that are estimated from  higher symmetry orbital corrections and
consistency of results carried out with different basis sets. They range from 0.2\% to 1.4\%. We note that these are only numerical uncertainties and
do not include estimates of missing theory (such as other triple and higher-excitation contributions). The uncertainties in our values represent our
best estimate of all possible sources of uncertainties, i.e. they give estimated boundary values of the recommended results. We refer the reader to
recent review \cite{JPBreview} for further discussion of the differences between numerical and complete uncertainties.

The same very large basis is used in all calculations carried out in this work. We use 70 basis set functions  for all partial waves with
$l_{max}\leq 6$. Use of such a large basis set results in negligible numerical errors in our values. The contribution of the $l>6$ partial waves to
the $4s-4p$ transitions is expected to be at the 0.05\% level. The contribution of the $l>6$ partial waves to the $3d-4p$ transitions is accounted
for by the scaling procedure. The correction is small since the entire scaling of the \textit{ab initio} SDpT values (that also accounts for the
corrections due to higher-order and non-linear excitations) is 0.7\%.

The other major differences between the present work and Ref.~\cite{das-09} include treatment of non-linear contributions, triple excitations, and
higher-excitation terms. The Breit correction calculated in \cite{das-09} is negligible for the E1 matrix elements. Ref.~\cite{das-09} include
non-liner terms at the SD level. While we have not explicitly included non-linear terms in this work, they were estimated by adjustment of the
correlation potential  described above along with contributions from higher excitations. We have demonstrated in Ref.~\cite{quad} that correcting
correlation potential staring from either linearized single-double coupled-cluster (LCCSD) or CCSD approximation leads to the same results within the
expected accuracy of the calculations. Our \textit{ab initio} inclusion of the valence triple excitation (SDpT) is more complete than that of
\cite{das-09} since we included triple corrections to both $\delta E_v$ and $\rho_{mv}$ equations while only $\delta E_v$ was corrected in
\cite{das-09}. We also estimated other higher-excitation corrections as describe above while no such estimates were done in Ref.~\cite{das-09}.

Our values are in agreement with results of \cite{das-09} for the $4p_{1/2}-4s$ and   $4p_{3/2}-3d_{3/2}$ transitions but disagree well beyond the
uncertainties for the $4p_{3/2}-4s$, $4p_{1/2}-3d_{3/2}$, and $4p_{3/2}-3d_{5/2}$ transitions. We find this to be rather irregular since the
correlation corrections are known to contribute nearly the same relative amount for the different  fine-structure transitions for such  light ions as
Ca$^+$.

 To clarify this issue, we calculated  the ratios $R=d_1^2/d^2_2$ of the squares of all three relevant pairs of matrix elements in the lowest-,
 second-, and third-order of MBPT, and all all-order approximations used in this work. The values of the matrix elements used in calculating the ratios
are listed in Table~\ref{tab-dip}. The comparison of all values is given in Table~\ref{comp2}. As we expected, all ratios of our values calculated in
all approximations, including the lowest-order DF values, are nearly identical to the non-relativistic values (2, 5, and 9, respectively) which are
simply the ratios of the corresponding angular factors. The effect of the entire correlation correction to the ratio is negligible. We see no
feasible explanation of the anomalous ratios in ~\cite{das-09} and significant changes between their GTO's/STO's and final value ratios. In the case
of the $4s-4p$ transitions, ratio of the final values from \cite{das-09} is 3$\sigma$ away from the NR value. The values of the other ratios are only
slightly outside of the numerical error but it is not clear what could cause such changes in ratios from initial values to the final ones. The
contributions from the higher symmetry orbitals can not change these ratios when the same basis set is used for the $nl$ states with different $j$.

In summary, we expect our results for all of the properties listed  in this work to be more accurate than that of Ref.~\cite{das-09} based on the
detailed analysis above.

\subsection{Transition rates and oscillator strengths}

We combine recommended NIST energies \cite{nist-web} and our final values of the matrix elements listed in Table~\ref{tab-dip} to calculate
transition rates $A$ and oscillator strengths $f$. The transition rates are calculated using
\begin{equation}
A_{ab}=\frac{2.02613\times10^{18}}{\lambda^3}\frac{S}{2j_a+1}\text{s}^{-1},
\end{equation}
where the wavelength $\lambda$ is in \AA~ and the line strength $S=d^2$ is in atomic units.

Transition rates $A$ (s$^{-1}$) and oscillator strengths ($f$) for the  55 $np-n^{\prime}s$, $np-n^{\prime}d$, and
 $nd-n^{\prime}f$ transitions in Ca~II
 are summarized in
Table~\ref{tab-osc1}. Vacuum wavelengths obtained from NIST energies are also listed for reference.  The relative uncertainties of the transition
rates and oscillator strengths are twice of  the corresponding matrix element uncertainties since these properties are proportional to the squares of
the matrix elements. The uncertainties in per cent are listed in the column labeled ``Unc.''.

\begin{table}
\caption{\label{tab-osc-av-comp} Comparison of the $j$-averaged oscillator strengths  with theoretical results from
Refs.~\protect\cite{life-th-89,mitroy-08}}
\begin{ruledtabular}
\begin{tabular}{lrrr}
\multicolumn{1}{c}{Transitions} & \multicolumn{1}{c}{Present}& \multicolumn{1}{c}{Ref.~\cite{life-th-89}} &
\multicolumn{1}{c}{Ref.~\cite{mitroy-08}} \\
\hline
 $4s-4p$&       0.970(9)       &0.9523   &0.9606 \\
 $4s-5p$&     1.25[-3]         &3.3[-4]  &1.72[-3]\\
 $5s-5p$&       1.486(5)       &1.4736   &        \\
 $4p-5s$&       0.178(2)       &0.168    &        \\
 $4p-4d$&       0.884(10)      &0.8741   &0.8685  \\
 $4p-5d$&       0.107(3)       &0.1076   &         \\
 $3d-4p$&       0.0645(7)      &0.0572   &0.0660   \\
 $3d-5p$&       5.1(1.8)[-4]   &6.9[-4]  & 3.8[-4]\\
 $4d-5p$&       0.188(3)       &0.1865   &        \\
 $3d-4f$&       0.154(3)       &         & 0.1599\\
 $4d-4f$&       1.098(9)       &         &       \\
\end{tabular}
\end{ruledtabular}
\end{table}

The values of the $j$-averaged oscillator strengths obtained using our final values of the matrix elements and NIST energies
 are compared with theoretical results from Refs.~\cite{life-th-89,mitroy-08} in
Table~\ref{tab-osc-av-comp}. The values of the Ref.~\cite{life-th-89} are obtained with semi-empirical approach that uses experimental energy levels
and experimental or theoretical core polarizabilities as an input and approximates the core potential by the Hartree-Slater method. In recent work,
Mitroy and Zhang \cite{mitroy-08} used non-relativistic configuration interaction with a semi-empirical core potential (CICP)  approach. The CICP
values are in good agreement with our results taking into account the accuracy of both calculations. Earlier and significantly less sophisticated
semi-empirical calculations of \cite{life-th-89} appear to be less accurate as expected.

\subsection{ Lifetimes and branching ratios}

We calculated lifetimes of the $5s$, $6s$, $7s$, $4p_j$, $5p_j$, $6p_j$, $4d_j$, $5d_j$, and $4f_j$  states in Ca$^+$ using out final values of the
dipole matrix elements and NIST energies \cite{nist-web}. The uncertainties in the lifetime values  are obtained from the uncertainties in the matrix
elements listed in Table~\ref{tab-dip}.
 The present
values are
  compared   with
available experimental ~\cite{life-expt-70,life-expt-75,life-4p-expt-93,life-4p-expt-94} and theoretical \cite{life-th-89,das-09} results in
Table~\ref{tab-life}.

The value of branching ratio of the $4p_{3/2}$ decay of Ca$^+$ calculated from transition rates given in Table~\ref{tab-osc1}
$A(4p_{3/2}-4s)$/$\sum_{j} A(4p_{3/2}-3d_{j})$=14.15(20) agrees  with 2008 measurement, 14.31(5), reported  in Ref.~\cite{branch-08} within our
uncertainty. As we noted above, our values for the $4s-4p$ and $4p-3d_{3/2}$ matrix elements are the same as in Ref.~\cite{mar-bbr-07} since the same
method is used. Therefore, the agreement of our values for the three branching fractions measured in \cite{branch-08} remains the same as listed in
the experimental work \cite{branch-08}: $R(4p_{3/2}-4s)=0.9347(3)^{\text{expt}}~\text{vs.}~0.9340^{\text{th}}$,
$R(4p_{3/2}-3d_{3/2})=0.00661(4)^{\text{expt}}~\text{vs.}~0.00667^{\text{th}}$,
$R(4p_{3/2}-3d_{3/2})=0.00587(2)^{\text{expt}}~\text{vs.}~0.00593^{\text{th}}$. The uncertainties in our values of the transition rates are about 1\%
for all three transitions. Therefore, the agreement of our central values with experiment  is significantly better than expected from our uncertainty
estimates (the uncertainty in the ratio is about twice that of the uncertainties in the individual transition rates). In fact, the $4s-4p_{3/2}$
branching fraction agrees with experiment to 0.07\% making substantial (3\%!) disagreement of our $4p$ lifetime values with 1993 experiment that
lists 0.3\% accuracy even more puzzling. Our calculation of the $4p$ lifetimes in K \cite{mar-pol-99} agrees with experimental values \cite{volz} to
0.13\%. Moreover, our primary $np_j$ lifetime values agree with recent experiments for all other alkali~\cite{mar-pol-99}, Sr$^+$ \cite{sr}, and
Ba$^{+}$\cite{ba}. Our values for the ground state polarizabilities that are completely dominated by the primary $ns-np$ matrix element
contributions, agree with all recent experiments in Li, Na, Cs~\cite{JPBreview}, Mg$^+$~\cite{mg}, Si$^{3+}$, and Ba$^+$\cite{ba}.
 This issue already
have been discussed in detail in both \cite{mar-bbr-07} and \cite{branch-08}. It would be very interesting to see new measurement of the $4p_j$
lifetimes, $4s-4p_j$ transition rates,  ground state polarizability, or other properties that allow to infer $4s-4p_j$ matrix elements in Ca$^+$.

All other experimental values listed in Table~\ref{tab-life} are much older (1970-1975) measurements with low
precision.
\begin{table} [t]
\caption{\label{tab-life} Comparison of the lifetimes (in nsec) of $nl_j$ states with other theory and experiment. Uncertainties are given in
parenthesis. References are given in square brackets.}
\begin{ruledtabular}
\begin{tabular}{llll}
\multicolumn{1}{c}{Level} & \multicolumn{1}{c}{Present}& \multicolumn{1}{c}{Expt.} &\multicolumn{1}{c}{Theory}
 \\
\hline
  $ 5s_{1/2}$&       3.91(4)&     4.3(4) \protect\cite{life-expt-70}  &4.153~\protect\cite{life-th-89}\\
  $ 6s_{1/2}$&       6.39(2)&                                              &6.766~\protect\cite{life-th-89}\\
  $ 7s_{1/2}$&      10.63(3)&                                              &11.262~\protect\cite{life-th-89}\\[0.4pc]
  $ 4p_{1/2}$&       6.88(6)&     7.098(20) \protect\cite{life-4p-expt-93}    &6.978(56) \protect\cite{das-09}\\
  $ 4p_{3/2}$&       6.69(6)&     6.924(19) \protect\cite{life-4p-expt-93}    &6.926(36) \protect\cite{das-09}\\
  $ 5p_{1/2}$&        35.4(7)&                                              & 36.200 \protect\cite{life-th-89}\\
  $ 5p_{3/2}$&        34.8(7)&                                              & 35.249 \protect\cite{life-th-89}\\
  $ 6p_{1/2}$&          89(2)&                                              & 100.254 \protect\cite{life-th-89}\\
  $ 6p_{3/2}$&          90(2)&                                              & 99.675  \protect\cite{life-th-89}\\[0.4pc]
  $ 4d_{3/2}$&        2.83(3)&     2.9(3) \protect\cite{life-expt-75}    &2.868 \protect\cite{life-th-89}\\
  $ 4d_{5/2}$&        2.85(3)&     3.1(2) \protect\cite{life-expt-70}    &2.886 \protect\cite{life-th-89}\\
  $ 5d_{3/2}$&       6.16(13)&     4.3(2) \protect\cite{life-expt-70}    &6.148 \protect\cite{life-th-89}\\
  $ 5d_{5/2}$&       6.21(14)&                                                &6.199 \protect\cite{life-th-89}\\[0.4pc]
  $ 4f_{5/2}$&        3.55(7)&                                               &3.895 \protect\cite{life-th-89}\\
  $ 4f_{7/2}$&        3.54(7)&                                               &3.897 \protect\cite{life-th-89}\\
\end{tabular}
\end{ruledtabular}
\end{table}
The values of the   metastable $3d_j$ state lifetimes calculated with our approach agree within the uncertainties with the  recent experimental
values \cite{ca-3d-expt-05}. This calculation and comparison with experiment was already discussed in  detail in \cite{ca-3d-expt-05}, and we do not
repeat it here. We note that 1\% RCC theoretical value for the $3d_{5/2}$ lifetime, $1.110(9)$s \cite{nd-das-06}, calculated by the same group as
work \cite{das-09} that we discussed at length in the matrix element section disagrees with both our value and experiment by 6\%. Nevertheless, their
$3d_{3/2}$ lifetime is in agrement with both our theoretical and experimental values. This demonstrates another significant inconsistency of the
approach used in \cite{das-09,nd-das-06} in calculations of properties of the levels from the same fine-structure multiplet. The ratio of these
lifetimes is affected very weakly by the correlation as discussed in detail in\cite{ca-3d-expt-05}. Our value of this ratio is 1.0259(9)
\cite{nd-das-06}, while the ratio between Ref.~\cite{nd-das-06} $3d$ lifetimes  is 1.068. We note that lowest-order DF ratio of these lifetimes is
1.0245.  Therefore, no difference in the treatment on the correlation correction can explain such anomalous ratio of these lifetimes.

%********************************
\section{ Static ground-state multipole polarizabilities of Ca~II}

The static multipole polarizability $\alpha^{Ek}$ of Ca$^+$ in its $4s$ ground state can be separated to a valence polarizability and a
polarizability of an ionic core.
 For the $4s$ state, the dominant valence
contribution  is calculated using the sum-over-state approach
\begin{equation}
\alpha^{Ek}_v = \frac{1}{2k+1}\sum_{nlj} \frac{|\langle nl_j\| r^k C_{kq} \|4s\rangle|^2}{E_{nl_j}-E_{4s}}, \label{alpha0}
\end{equation}
where  $C_{kq}(\hat{r})$ is a normalized spherical harmonic and $nl_j$ is $np_{j}$, $nd_{j}$, and $nf_j$ for $k$ = 1, 2, and 3, respectively
\cite{mg}. The E2 and E3 matrix elements and their uncertainties are calculated  following the same approach that we used in calculating
electric-dipole matrix elements (see Section~\ref{E1}).

 Contributions to the  ground-state state dipole, quadrupole, and octupole polarizabilities
are presented in Table~\ref{tab-pol-mult}. Dominant contributions are listed separately. The remainders of the sums are listed together. For example,
row labeled ``$nd_{3/2}$'' gives the combined contribution of all $nd_{3/2}$ terms with $n > 7$. The first terms ($4p$, $3d$, and $4f$, respectively)
in the sum-over-states for $\alpha^{E1}$, $\alpha^{E2}$, and $\alpha^{E3}$ contribute 99.7\%, 59\%, and 79\%, respectively, of the total valence
polarizabilities. The rapid convergence of the sum over states for $\alpha^{E1}$ has been emphasized in many publications (for example,
Refs.~\cite{mar-pol-99,pol-der-99}). The sums in Eq.~(\ref{alpha0}) converge much slower for the $E2$ and $E3$ polarizabilities. Therefore, accurate
evaluation of a large number of terms in the sums~(\ref{alpha0}) is needed for these states.
 We use NIST energies from \cite{nist-web} and our  final recommended values of the matrix elements to evaluate terms with $n \leq 13$. We use theoretical SD energies and matrix elements  to evaluate terms with $13\leq n \leq 26$. The remaining contributions to
$\alpha^{Ek}$ from basis functions with $27 \leq n \leq 70$ are evaluated in the DF approximation.  These remainders  are very small. Even in the
case of the E3 polarizability, which is the slowest one to converge, the tail remainder with $n > 26$ contributes only 13~~a.u. which is 0.14\% of
the total valence polarizability.

The electric-dipole core polarizability is taken to be 3.26(3)~a.u. based on the comparison of the coupled-cluster and experimental values listed in
the review~\cite{JPBreview}. This value is essentially the same as the random-phase approximation result of  3.25~a.u. Since this value is the
polarizability of the ionic Ca$^{2+}$ core, we need to account for the presence of the valence electron by adding a term $\alpha_{vc}$ which in this
case is equal to half of the core polarizability contribution from the excitation to the valence $4s$ shell.   In the cases of the E2 and E3
polarizabilities, we evaluate core contributions in the random-phase approximation \cite{RPA}. The core polarizabilities are small in comparison with
the valence ones and their uncertainties are negligible. We note that $\alpha_{vc}$ terms are zero for the E2 and E3 polarizabilities since Ca$^+$
core contains no $nd$ or $nf$ states.

Our final results for the ground-state multipole polarizabilities  are compared  with other theoretical values \cite{mitroy-08,das-09} in
Table~\ref{tab-pol-mult}. The CICP values of Ref.~\cite{mitroy-08} are in remarkably good agreement with our results in all three cases.
Interestingly, we differ by 3\% in the main $4s-3d_j$ contribution to the E2 polarizabilities (which is 523~a.u. in \cite{mitroy-08}), but agree in
the final value. The difference between the present and Ref.~\cite{das-09} E1 polarizability value  results from the differences in the $4s-4p_j$
matrix elements
 which we already discussed in detail in Section~\ref{E1}.

\begin{table}
\caption{ Contributions to dipole, $\alpha^{E1}$, quadrupole, $\alpha^{E2}$, and  octupole, $\alpha^{E3}$, polarizabilities (a.u.) of the Ca$^+$
ground states. All values are in atomic units. The uncertainties are given in parenthesis. \label{tab-pol-mult} }.
\begin{ruledtabular}
\begin{tabular}{lrlrlr}
 \multicolumn{2}{c}{$\alpha^{E1}$} &
 \multicolumn{2}{c}{$\alpha^{E2}$}&
\multicolumn{2}{c}{$\alpha^{E3}$}
 \\
 \hline
$     4p_{1/2}      $   &   24.4(2) &   $     3d_{3/2}    $   &   203(2)  &   $     4f_{5/2}  $   &   3017(31)    \\
$     4p_{3/2}      $   &   48.4(4) &   $     4d_{3/2}    $   &   125(1)  &   $     5f_{5/2}  $   &   469(4)  \\
$   np_{1/2}        $   &   0.08(2) &   $     5d_{3/2}    $   &   10.6(2) &   $     6f_{5/2}  $   &   141(1)  \\
$   np_{3/2}        $   &   0.14(3) &   $     6d_{3/2}    $   &   2.7(1)  &   $     7f_{5/2}  $   &   59.5(4) \\
    Core                &   3.26(3) &   $     7d_{3/2}    $   &   1.1(0)  &   $     8f_{5/2}  $   &   30.7(1) \\
    $\alpha_{\text{vc}}$&   -0.12(1)&   $     nd_{3/2}   $   &   4.0(2)  &   $      nf_{5/2}   $   &   131(19) \\
    Total               &   76.1(5) &   $     3d_{5/2}   $   &   304(3)  &   $     4f_{7/2} $   &   4023(41)    \\
& &                                     $     4d_{5/2}   $   &   187(2)  &   $     5f_{7/2} $   &   625(5)  \\
                        &           &   $     5d_{5/2}   $   &   15.9(2) &   $     6f_{7/2} $   &   188(1)  \\
                        &           &   $     6d_{5/2}   $   &   4.1(1)  &   $     7f_{7/2} $   &   79.3(5) \\
                        &           &   $     7d_{5/2}   $   &   1.6(0)  &   $     8f_{7/2} $   &   41.0(1) \\
                        &           &   $     nd_{5/2}   $   &   5.0(3)  &   $     nf_{7/2} $   &   175(24) \\
                        &           &       Core        &   6.9(3)  &       Core        &   34(4)   \\
       &          &       Total       &   871(4)  &       Total       &   9012(60)    \\[0.3pc]
 Ref.~\protect\cite{mitroy-08}& 75.49 &Ref.~\protect\cite{mitroy-08}&875.1&Ref.~\protect\cite{mitroy-08}&8990\\
 Ref.~\protect\cite{das-09}        &73.0(1.5)& &&&\\
\end{tabular}
\end{ruledtabular}
\end{table}

\begin{table*}
\caption{\label{tab-scalar} Contributions to the $5s$, $6s$, $7s$, $8s$, $4p_j$, $5p_j$, $3d_j$, and $4d_j$ scalar polarizabilities of Ca~II in
$a_0^3$. Uncertainties are given in parenthesis. The final results are compared with other theory \cite{das-09,mitroy-08}. \label{pol1}}
\begin{ruledtabular}
\begin{tabular}{lrlrlrlr}
 \multicolumn{1}{c}{Contribution} &  \multicolumn{1}{c}{$\alpha_0$} &\multicolumn{1}{c}{Contribution} &
 \multicolumn{1}{c}{$\alpha_0$}&
 \multicolumn{1}{c}{Contribution} &  \multicolumn{1}{c}{$\alpha_0$}  &\multicolumn{1}{c}{Contribution} &  \multicolumn{1}{c}{$\alpha_0$}
 \\
 \hline
\multicolumn{2}{c}{${5s}$}& \multicolumn{2}{c}{${6s}$} & \multicolumn{2}{c}{${7s}$}& \multicolumn{2}{c}{${8s}$}
 \\ [0.3pc]
$  4p_{1/2}$&   -11.7(1)    &   $  5p_{1/2}$   &-137.1(5)  &   $   6p_{1/2}  $&   -799(1)  &$   7p_{1/2}  $   &   -3214(4)    \\
$  4p_{3/2}$&   -24.0(2)    &   $  5p_{3/2}$   &   -282(1) &   $   6p_{3/2}  $&   -1639(2) &$   7p_{3/2}  $   &   -6593(8)    \\
$  5p_{1/2}$&   339(1)      &   $  6p_{1/2}$   &   2118(5) &   $   7p_{1/2}  $&   8894(21) &$   8p_{1/2}  $   &   29153(125)  \\
$  5p_{3/2}$&   671(2)      &   $  6p_{3/2}$   &   4180(10)&   $   7p_{3/2}  $&   17532(42)&$   8p_{3/2}  $   &   57432(248)  \\
    Other       &   3.6 &       Other       &   3.8 &       Other       &   5   &       Other       &   -22 \\
    Total       &   978(3)  &       Total       &   5882(11)    &       Total       &   23990(50)   &       Total       &   76760(280)  \\
    Ref.~\cite{mitroy-08} & 983.5 &           &                &                   &               &                   & \\[0.3pc]
\multicolumn{2}{c}{{$4p_{1/2}$}}& \multicolumn{2}{c}{$5p_{1/2}$} & \multicolumn{2}{c}{$4d_{3/2}$}& \multicolumn{2}{c}{$4d_{5/2}$} \\[0.3pc]
$  4s     $   &   -24.39(22)  &   $   5s      $   &   -339(1) &   $  4p_{1/2}   $   &   -21.2(3)    &   $   4p_{3/2}   $   &   -25.9(3)    \\
$   5s    $   &   11.66(12)   &   $   6s      $   &   137     &   $  5p_{1/2}   $   &   547(9)      &   $   5p_{3/2}   $   &   645(10) \\
$   6s    $   &   0.54        &   $   7s      $   &   5       &   $  4p_{3/2}   $   &   -4.3        &   $   np_{3/2}   $   &   0   \\
$ ns      $   &   0.46(1)     &   $   ns      $   &   3       &   $  5p_{3/2}   $   &   107(2)      &   $   4f_{5/2}   $   &   20.1(2) \\
$ 3d_{3/2}    $   &   -38.47(41)  &   $    4d_{3/2}    $   &   -1094(17)&  $   np      $   &   0           &   $   nf_{5/2}   $   &   0.1 \\
$ 4d_{3/2}    $   &   42.38(50)   &   $    5d_{3/2}    $   &   337(5)  &   $   4f_{5/2}   $   &   421(4)      &   $ 4f_{7/2}   $   &   402(3)  \\
$ 5d_{3/2}    $   &   2.29(6)     &   $    6d_{3/2}    $   &   18      &   $   5f_{5/2}   $   &   0.1         &   $ 5f_{7/2}   $   &   0.1 \\
$ nd_{3/2}    $   &   1.51(8)     &   $    nd_{3/2}    $   &   11(1)   &   $   nf_{5/2}   $   &   2.3(1)      &   $ nf_{7/2}   $   &   2.2(1)  \\
    Core        &   3.26(3) &       Core        &   3   &       Core      &   3.3 &       Core        &   3.3 \\
    Total       &   -0.75(70)   &       Total       &   -920(18)    &       Total       &   1054(10)    &       Total       &   1046(11)    \\
   Ref.~\cite{mitroy-08}& -2.032 &       Ref.~\cite{mitroy-08}& -1135 &   Ref.~\cite{mitroy-08}&       1209 &   Ref.~\cite{mitroy-08}& 1209 \\   [0.3pc]
\multicolumn{2}{c}{$4p_{3/2}$}& \multicolumn{2}{c}{$5p_{3/2}$} & \multicolumn{2}{c}{$3d_{3/2}$}& \multicolumn{2}{c}{$3d_{5/2}$} \\[0.3pc]
$  4s  $   &   -24.18(22)  &   $   5s      $   &   -335(1) &$    4p_{1/2}   $   &   19.24(21)   &$  4p_{3/2}   $   &   22.78(25)   \\
$  5s  $   &   12.02(11)   &   $   6s      $   &   141 &   $     np_{1/2}   $   &   0.02    &   $   np_{3/2}   $   &   0.03    \\
$  6s  $   &   0.55    &       $   7s      $   &   5   &   $     4p_{3/2}   $   &   3.76(4) &   $   4f_{5/2}   $   &   0.12    \\
$  ns  $   &   0.46(1) &       $   ns      $   &   3   &   $     np_{3/2}   $   &   0.01    &   $   nf_{5/2}   $   &   0.17    \\
$  3d_{3/2}   $   &   -3.76(4) &  $   4d_{3/2}   $   &   -107(2) & $    4f_{5/2}   $   &   2.50(6) &   $    4f_{7/2}   $   &   2.39(5) \\
$  4d_{3/2}   $   &   4.33(5) &   $   5d_{3/2}   $   &   34(1)   &$     5f_{5/2}   $   &   0.81(1) &   $    5f_{7/2}   $   &   0.77(1) \\
$  5d_{3/2}   $   &   0.23(1) &   $   6d_{3/2}   $   &   2   &   $      6f_{5/2}   $   &   0.37(1) &   $    6f_{7/2}   $   &   0.35(1) \\
$  nd_{3/2}   $   &   0.15(1) &   $   nd_{3/2}   $   &   1   &   $      7f_{5/2}   $   &   0.20(1) &   $    7f_{7/2}   $   &   0.19(1) \\
$  3d_{5/2}   $   &   -34.17(37)&$  4d_{5/2}   $   &   -967(16) &$    (8-12)f_{5/2}  $   &   0.33(1) &   $   (8-12)f_{7/2}  $   &   0.31(1) \\
$  4d_{5/2}   $   &   38.85(45)& $  5d_{5/2}   $   &   309(5)  &$     (13-26)f_{5/2} $   &   1.42(4) &   $   (13-26)f_{7/2} $   &   1.39(4) \\
$  5d_{5/2}   $   &   2.07(5) &  $  6d_{5/2}   $   &   16  &   $      nf_{5/2}   $   &   0.32(19)    &   $   nf_{7/2}   $   &   0.27(15)    \\
$  nd_{5/2}   $   &   1.22(6) &  $  nd_{5/2}   $   &   9   &       Core        &   3.26(3) &       Core        &   3.26(3) \\
    Core        &   3.26(3) &       Core        &   3   &       $\alpha_{vc}$       &   -0.23(1)    &       $\alpha_{vc}$       &   -0.23(1)    \\
    Total       &   1.02(64)    &       Total       &   -886(16)    &       Total       &   32.0(3) &       Total       &   31.8(3) \\
   Ref.~\cite{mitroy-08}& -2.032 &       Ref.~\cite{mitroy-08}& -1135 &  Ref.~\cite{mitroy-08} & 32.73 &Ref.~\cite{mitroy-08} & 32.73 \\
                   &       &                   &                 &  Ref.~\cite{das-09}&28.5(1.0) &  Ref.~\cite{das-09}& 29.5(1.0)\\
\end{tabular}
\end{ruledtabular}
\end{table*}

\begin{table}
\caption{Contributions to the $4p_{3/2}$, $5p_{3/2}$, $3d_j$, and $4d_j$ tensor polarizabilities of Ca~II in $a_0^3$. Uncertainties are given in
parenthesis. The final results are compared with other theory \cite{das-09,mitroy-08}. \label{pol2} }
\begin{ruledtabular}
\begin{tabular}{lrlr}
 \multicolumn{1}{c}{Contribution} &  \multicolumn{1}{c}{$\alpha_2$} &\multicolumn{1}{c}{Contribution} &
 \multicolumn{1}{c}{$\alpha_2$}
 \\
 \hline
\multicolumn{2}{c}{$4p_{3/2}$}& \multicolumn{2}{c}{$5p_{3/2}$}\\ [0.3pc]
$   4s  $   &   24.18(22)   &   $   5s  $   &   335(1)  \\
$   5s  $   &   -12.02(11)  &   $   6s  $   &   -141    \\
$   6s  $   &   -0.55   &   $   7s  $   &   -5  \\
$   ns  $   &   -0.46(1)    &   $   ns  $   &   -3  \\
$   3d_{3/2}   $   &   -3.01(3)    &   $   4d_{3/2}   $   &   -85(1)  \\
$   4d_{3/2}   $   &   3.46(4) &       $   5d_{3/2}   $   &   28  \\
$   5d_{3/2}   $   &   0.18    &       $   6d_{3/2}   $   &   1   \\
$   nd_{3/2}   $   &   0.12    &       $   nd_{3/2}   $   &   1   \\
$   3d_{5/2}   $   &   6.83(7) &       $   4d_{5/2}   $   &   193(3)  \\
$   4d_{5/2}   $   &   -7.77(9)    &   $   5d_{5/2}   $   &   -62(1)  \\
$   5d_{5/2}   $   &   -0.41(1)    &   $   6d_{5/2}   $   &   -3   \\
$   nd_{5/2}   $   &   -0.24(1)    &   $   nd_{5/2}   $   &   -2  \\
    Total       &   10.31(28)   &       Total       &   258(4)  \\
Ref.~\cite{mitroy-08}& 10.47 &  Ref.~\cite{mitroy-08}& 286.2 \\
\multicolumn{2}{c}{$3d_{3/2}$}& \multicolumn{2}{c}{$3d_{5/2}$}\\ [0.3pc]
$   4p_{1/2}   $   &   -19.24(21)  &   $   4p_{3/2}   $   &   -22.78(25)  \\
$   np_{1/2}   $   &   -0.01   &       $   np_{3/2}   $   &   -0.02   \\
$   4p_{3/2}   $   &   3.01(3) &       $   4f_{5/2}   $   &   0.18    \\
$   np_{3/2}   $   &   0.00    &       $   nf_{5/2}   $   &   0.15    \\
$   4f_{5/2}   $   &   -0.50(1)    &   $   4f_{7/2}   $   &   -0.85(2)    \\
$   5f_{5/2}   $   &   -0.16   &       $   5f_{7/2}   $   &   -0.28   \\
$   6f_{5/2}   $   &   -0.07   &       $   6f_{7/2}   $   &   -0.12   \\
$   nf_{5/2}   $   &   -0.46(2)    &   $   nf_{7/2}   $   &   -0.79(4)    \\
    Total       &   -17.43(23)  &       Total       &   -24.51(29)  \\
     Ref.~\cite{das-09}& -15.8(7) &  Ref.~\cite{das-09}& -22.45(5) \\
Ref.~\cite{mitroy-08}& -17.64&  Ref.~\cite{mitroy-08}& -25.20 \\
 \multicolumn{2}{c}{$4d_{3/2}$}&
\multicolumn{2}{c}{$4d_{5/2}$}\\ [0.3pc]
$   4p_{1/2}   $   &   21.2(3) &   $   4p_{3/2}   $   &   25.9(3) \\
$   5p_{1/2}   $   &   -547(9) &   $   5p_{3/2}   $   &   -645(10)    \\
$   4p_{3/2}   $   &   -3.46(4) &  $   np_{3/2}   $   &   -0.02   \\
$   5p_{3/2}   $   &   85(1)   &   $   4f_{5/2}   $   &   23.0(2) \\
$   np        $   &   -0.01   &    $   nf_{5/2}   $   &   0.13    \\
$   4f_{5/2}   $   &   -84.1(7) &  $   4f_{7/2}   $   &   -143(1) \\
$   5f_{5/2}   $   &   -0.03(1) &  $   5f_{7/2}   $   &   -0.05(2)    \\
$   nf_{5/2}   $   &   -0.46   &   $   nf_{7/2}   $   &   -0.79   \\
    Total       &   -529(9) &       Total       &   -740(10)    \\
    Ref.~\cite{mitroy-08}& -615.9&  Ref.~\cite{mitroy-08}& -879.8 \\
\end{tabular}
\end{ruledtabular}
\end{table}

\section{ Scalar  and tensor excited
state polarizabilities}

The valence scalar $\alpha_{0}(v)$  and tensor $\alpha_{2}$ polarizabilities of Ca$^+$  in an excited state $v$  are given by
%***********************
\begin{equation}
\alpha _{\text{0}}(v)\ =\frac{2}{3(2j_{v}+1)}\sum_{nlj}\frac{|\langle v||d||nlj\rangle |^{2}}{E_{nlj}-E_{v}}, \label{eqp1}
\end{equation}

\begin{align}
\alpha _{2}& \ =(-1)^{j_{v}}\sqrt{\frac{40j_{v}(2j_{v}-1)}{%
3(j_{v}+1)(2j_{v}+1)(2j_{v}+3)}}\   \nonumber  \label{eq1a} \\
& \times \sum_{nlj}(-1)^{j}\left\{
\begin{array}{lll}
j_{v} & 1 & j \\
1 & j_{v} & 2
\end{array}
\right\} \frac{|\langle v||d||nlj\rangle |^{2}}{E_{nlj}-E_{v}} \ .
\end{align}
The ionic core polarizability discussed in the previous section has to be added to the valence term given by Eq.~(\ref{eqp1}) and corrected for the
presence of the respective valence electron (term $\alpha_{vc}$). This core correction  $\alpha_{vc}$ term is negligible for all excited states that
we considered with the exception of the $3d_{3/2}$ and $3d_{5/2}$ states. It is calculated in the RPA.

The dipole polarizability calculations are carried out in the same way as the calculations of the multipole polarizabilities discussed in the
previous section. We list the contributions to the   $5s$, $6s$, $7s$, $8s$, $4p_j$, $5p_j$, $3d_j$, and $4d_j$ scalar polarizabilities of Ca$^+$
 in Table~\ref{pol1}. The dominant contributions are listed
separately. The remaining contributions are grouped together. For example, ``$nd_{3/2}$'' contribution includes all of the $nd_{3/2}$ terms excluding
only the terms that were already listed separately.

The Table~\ref{pol1} illustrates very fast convergence of the $ns$ level polarizabilities which are dominated by the corresponding $np$
contributions. The $(n-1)p$ term contributions are significant, while all of the other contributions are very small owing to fast convergence of the
sums. We use our recommended values for the $7p$, and $8p$ energy levels, for which we did not find the experimental values. The uncertainties in
these recommended energy values are included when the polarizability uncertainties are calculated. The uncertainties of the final polarizability
values are obtained by adding the uncertainties of the individual terms in quadarture.

When calculating polarizabilities of the $np$ and $nd$ states, we calculated the terms with $n\leq 26$ using the all-order approach. The terms with
$n>10$ are calculated using our calculated recommended values of the E1 matrix elements and the experimental energies ~\cite{nist-web}. The terms
with $10<n<27$ are calculated using SD energies and matrix elements. The remainders are evaluated in the DF or RPA approximations. The remainders in
the $3d-nf$ sums are treated more accurately as described below.

 We find that the scalar polarizabilities of the $4p_{3/2}$ and $4p_{1/2}$ states are anomalously small owing to a very precise cancelations of the
 various contributions. This fact was already pointed out by Mitroy and Zhang \cite{mitroy-08}. Our uncertainties of these polarizability values
  are very large because of these severe cancelations. Such cancelations are not observed for the $5p$ polarizabilities, where $5p-4d$
  contributions strongly dominate.

The case of the $3d_{5/2}$ polarizability is particularly interesting owing to its importance for the calculation of the blackbody radiation shift in
the optical frequency standard with Ca$^+$ ion. Ref.~\cite{mar-bbr-07} points out  that the sum over the $nf_{7/2}$ states converges very slowly
making accurate calculation of these contributions difficult. In this work, we have explored several different approaches to the accurate calculation
of this sum and obtained consistent results in all cases.

First, we calculated terms with  $n\leq 26$ in the all-order approximation, and determined that these terms contribute 5.41(16)~a.u. We find that
even with so many terms included, the remainder is still $0.43$~a.u. in DF approximation and $0.32$~a.u. in the RPA, which is a significant fraction
of the total $nf_{7/2}$ contribution. Therefore, we estimate the accuracy of the DF approximation by calculating the main terms with $n\leq 26$ in
the DF approximation as well. We find that  DF approximation overestimates the polarizability contributions from highly-exited state by about 38\%
and adjust the DF value accordingly. The entire adjustment is taken to be the uncertainly of the $n>26$ remainder. Therefore,  our final value for
$n>26$ $nf_{7/2}$ terms is 0.27(15), and the total $nf_{7/2}$ sum is equal to 5.67(22)~a.u.

Second, we carry out the calculation of the highly-excited states by a different approach to verify that the all-order calculation of such
highly-excited states does not introduce unexpected errors. We compare the contributions with $6<n<13$ calculated in  DF, RPA, and all-order
approximations.   The wave functions of all these states fit inside of our 220~a.u. cavity, and the all-order method is definitely reliable for these
states. We establish that DF overestimates the results by 40-44\% and RPA overestimates the results by 37-41\%. Then, we use these percentages to
adjust the contributions for the $n>12$ states calculated in the DF and RPA approximations. Adding these adjusted remainders to the all-order terms
with $n<13$ gives 5.63~a.u. (DF) and 5.53~a.u. (RPA) for the total $nf_{7/2}$ sum. These values are consistent with our result 5.67(22) obtained
above. Such accurate evaluation of this sum allows us to reduce the uncertainty of the $3d_{5/2}$ polarizability by a factor of 3 in comparison with
the previous calculation of this quantity using the all-order sum-over-states approach~\cite{mar-bbr-07}. The calculation of the $nf_{5/2}$
contribution to the $3d_{3/2}$ polarizability is carried out using the same method. In all the other cases, the contributions of the terms with
$n>26$ are very small in comparison with the other terms and DF approximation is sufficiently accurate.

We compare our values of the scalar polarizabilities for the $3d_{3/2}$ and  $3d_{5/2}$ with RCC  \cite{das-09} and CICP \cite{mitroy-08} theoretical
calculations in Table~\ref{pol1}. Ref.~\cite{das-09} estimates their numerical (basis set truncation) uncertainty to be 3.5\%. This estimate does not
account for the uncertainty owing to the missing correlation correction. Our $3d_j$ polarizabilities differ from values of \cite{das-09} by 7\% and
11\%. It is well known (see \cite{JPBreview} and references therein) that the polarizabilities are very sensitive to the problems with the basis set
completeness in the coupled-cluster calculations such as RCC calculation of Ref.~\cite{das-09}. We note that Ref.~\cite{das-09} STO values (31.6 and
32.5) are very close to our results.

 Since the CICP calculation \cite{mitroy-08} is non-relativistic, we list their values for
both fine-structure states. Our $5s$ and $3d$ results are in good agreement with CICP calculation. The agreement is rather poor for the $4p$, $4d$,
and $5p$ states. It is expected for the $4p$ states owing to severe cancelations discussed above but somewhat surprising for the other two states.

We list the contributions to the tensor polarizabilities of the Ca$^+$ in $4p_{3/2}$, $5p_{3/2}$, $3d_j$, and $4d_j$ states in Table~\ref{pol2}.
Tensor polarizability calculations are carried our in the same way as the scalar polarizability ones. The same designations are used in
Table~\ref{pol2} as in the scalar polarizability Table~\ref{pol1}. The final values are compared with RCC  \cite{das-09} and CICP \cite{mitroy-08}
theoretical calculations. We multiply the $3d$ and $4d$ non-relativistic values of \cite{mitroy-08} by $7/10$ to compare these values to our
$3d_{3/2}$ and $4d_{3/2}$ tensor polarizabilities (see \cite{JPBreview} for explanation of this conversion factor). The differences between the
present and \cite{mitroy-08,das-09} theoretical tensor polarizability values are similar to those for the scalar polarizabilities for these states.
\begin{table}
\caption{\label{tab-hyp} Hyperfine constants $A$ (in MHz)  in
 $^{43}$Ca$^+$ ($I=7/2$, $\mu=-1.31727$).
 The SD and SDpT  all-order
results are compared with theoretical~\cite{hyp-04} and experimental~\cite{hyp-94Z,hyp-expt-98} values.}
\begin{ruledtabular}
\begin{tabular}{lrrrrr}
\multicolumn{1}{c}{Level} & \multicolumn{1}{c}{$A^{\rm DF}$} & \multicolumn{1}{c}{$A^{\rm SD}$} & \multicolumn{1}{c}{$A^{\rm SDpT}$} &
\multicolumn{1}{c}{Th.~\cite{hyp-04}} &
\multicolumn{1}{c}{Expt.~\cite{hyp-94Z,hyp-expt-98}} \\
\hline
 $4s $ &  -587.39&   -818.82 & -801.28 &-805.348&-806.402072\\
 $5s $ &  -183.99&   -239.19 & -236.30 &&\\
 $6s $ &   -81.25&   -103.29 & -102.33 &&\\
 $7s $ &   -42.91&    -53.96 &  -53.53 &&\\[0.4pc]
 $4p_{1/2} $ &  -101.48&   -148.26 & -144.96 &-143.068&-145.4(1)\\
 $5p_{1/2} $ &   -37.43&    -51.28 &  -50.71 &&\\
 $6p_{1/2} $ &   -17.91&    -24.14 &  -23.81 &&\\
 $7p_{1/2} $ &    -9.94&    -13.25 &  -13.08 &&\\[0.4pc]
 $4p_{3/2} $ &   -19.64&    -31.04 &  -30.34 &-30.498&-31.0(2)\\
 $5p_{3/2} $ &    -7.25&    -10.69 &  -10.58 &&\\
 $6p_{3/2} $ &    -3.47&     -5.04 &   -4.97 &&\\
 $7p_{3/2} $ &    -1.93&     -2.76 &   -2.73 &&\\[0.4pc]
 $3d_{3/2} $ &  -33.20 &   -48.83  &  -47.63 &-47.824&-47.3(2)\\
 $4d_{3/2} $ &   -8.01 &    -9.32  &   -9.35 &&\\
 $5d_{3/2} $ &   -3.36 &    -3.86  &   -3.88 &&\\
 $6d_{3/2} $ &   -1.75 &    -1.99  &         &&\\
 $7d_{3/2} $ &   -1.03 &    -1.17  &         &\\[0.4pc]
 $3d_{5/2} $ &   -14.14&     -4.71 &  -4.24  &-3.553&-3.8(6)\\
 $4d_{5/2} $ &    -3.41&     -3.12 &  -3.06  &&\\
 $5d_{5/2} $ &    -1.43&     -1.51 &  -1.49  &&\\
 $6d_{5/2} $ &    -0.748&    -0.831&         &&\\
 $7d_{5/2} $ &    -0.441&    -0.502&         &&\\[0.4pc]
 $4f_{5/2} $ &    -0.151&    -0.165&  -0.163 &&\\
 $5f_{5/2} $ &    -0.079&    -0.089&         &&\\
 $6f_{5/2} $ &    -0.046&    -0.053&         &&\\
 $7f_{5/2} $ &    -0.029&    -0.034&         &&\\[0.4pc]
 $4f_{7/2} $ &    -0.084&    -0.044&  -0.044 &&\\
 $5f_{7/2} $ &    -0.044&    -0.011&         &&\\
 $6f_{7/2} $ &    -0.026&    -0.002&         &&\\
 $7f_{7/2} $ &    -0.016&     0.000&         &&\\
\end{tabular}
\end{ruledtabular}
\end{table}

\begin{table*}
\caption{\label{tab-hypb} Hyperfine constants $B$ (in MHz)  in
  $^{43}$Ca$^+$.
 Nuclear quadrupole moment $Q$ is taken to be
 equal to -0.044(9)~barns (1~b=10$^{-24}$cm$^2$) \cite{hyp-04}.
 The SD and SDpT values are compared with theory \protect\cite{hyp-04}  and experiment \protect\cite{hyp-expt-98}.}
\begin{ruledtabular}
\begin{tabular}{lrrrrrrrl}
\multicolumn{1}{c}{Level} & \multicolumn{1}{c}{$\frac{B^{\rm DF}}{Q}$} & \multicolumn{1}{c}{$\frac{B^{\rm SD}}{Q}$} &
\multicolumn{1}{c}{$\frac{B^{\rm SDpT}}{Q}$} & \multicolumn{1}{c}{$B^{\rm DF}$} & \multicolumn{1}{c}{$B^{\rm SD}$} & \multicolumn{1}{c}{$B^{\rm
SDpT}$} & \multicolumn{1}{c}{$\frac{B^{\rm th}}{Q}$~\protect\cite{hyp-04}  } &
\multicolumn{1}{c}{$B^{\rm expt}$~\protect\cite{hyp-expt-98}} \\
\hline
$4p_{3/2} $ &    96.69&   153.99&   150.81&  -4.25&     -6.78&     -6.64& -151.798  &-6.9(1.7)\\
$5p_{3/2} $ &    35.58&    52.40&    51.82&  -1.57&     -2.31&     -2.28&    &   \\
$6p_{3/2} $ &    16.93&    24.37&    24.06&  -0.75&     -1.07&     -1.06&    &   \\

$3d_{3/2} $ &    54.24&    67.01&    65.24&  -2.39&     -2.95&     -2.87& -68.067 &-3.7(1.9)     \\
$4d_{3/2} $ &    12.93&    17.37&    17.24&  -0.57&     -0.76&     -0.76&    &     \\
$5d_{3/2} $ &     5.31&     7.28&     7.24&  -0.23&     -0.32&     -0.32&    &     \\

$3d_{5/2} $ &    76.86&    95.19&    92.69&  -3.38&     -4.19&     -4.08& -100.208&-3.9(6.0)     \\
$4d_{5/2} $ &    18.33&    24.70&    24.51&  -0.81&     -1.09&     -1.08&    &     \\
$5d_{5/2} $ &     7.54&    10.35&    10.30&  -0.33&     -0.46&     -0.45&    &     \\
\end{tabular}
\end{ruledtabular}
\end{table*}

%***************
\section{Blackbody radiation shift in Ca$^+$ optical frequency standard}
The electrical field $E$ radiated by a blackbody at temperature $T$, as given by Planck's law,
\begin{equation}
E^2(\omega) d \omega=\frac{8 \alpha^3}{\pi} \frac{\omega^3 d\omega}{\mathrm{exp}(\omega/k_B T)-1},
\end{equation}
induces a nonresonant perturbation of the optical transition at room temperature~\cite{BBR_Farley}. The frequency shift of an atomic state due to
such an electrical field is related to the static electric-dipole polarizability $\alpha_0$ by
 (see Ref.~\cite{BBR_Porsev})
\begin{equation}
 \Delta \nu = -\frac{1}{2}(831.9~{\rm V/m})^2
\left( \frac{T(K)}{300} \right)^4 \alpha_0(1+\eta),
\end{equation}
where  $\eta$ is a small dynamic correction due to the frequency distribution. Only the electric-dipole transition part of the contribution is
considered in the formula above because the contributions from M1 and E2 transitions are suppressed by a factor of $\alpha^2$ \cite{BBR_Porsev}.  The
overall BBR shift of the Ca$^+$ $4s-3d_{5/2}$ clock transition frequency is then calculated as the difference between the BBR shifts of the
individual levels involved in the transition:
\begin{eqnarray} \label{eq:BBR}
 \Delta_{\mathrm{BBR}}(4s - 3d_{5/2})& =& -\frac{1}{2}
[\alpha_0(3d_{5/2})-\alpha_0(4s_{1/2})]  \nonumber \\
&\times & (831.9~{\rm V/m})^2 \left( \frac{T(K)}{300} \right)^4.
\end{eqnarray}
 The  tensor part of polarizability is averaged out due to the isotropic nature of the electric field radiated
by the blackbody. Substituting out values for the $4s$ and $3d_{5/2}$ static polarizabilities into Eq.~(\ref{eq:BBR}), we obtain 0.3815(44)~Hz for
the BBR shift.
 We note that atomic units for $\alpha$ are converted to SI units via $\alpha / h
[$Hz$/(V/m)^2]= 2.48832 \times 10 ^{-8} \alpha [a.u.]$, where the conversion coefficient is $4 \pi \epsilon_0 a_0^3 /h$ and Planck constant $h$ is
factored out.

We estimate the dynamic corrections  to be  $\eta$=0.0012
 and $\eta$=0.0044
 for the $4s$ and $3d_{3/2}$ states,
respectively, following Ref.~\cite{BBR_Porsev}. The resulting dynamic correction to the  BBR shift is $-0.0004$~Hz and our final value is
$$\Delta_{\mathrm{BBR}}(4s - 3d_{5/2})=0.3811(44)~\textrm{Hz}.$$
The value is the same for different Ca$^+$ isotopes within its accuracy.  The third-order $F$-dependent polarizability of the ground state is
evaluated in the last section of this paper. Its contribution is several orders of magnitudes smaller than the second-order value and can be omitted
in evaluating BBR shift in the optical standard.

The present value is consistent with other calculations, 0.380(14)~Hz \cite{mar-bbr-07}, 0.37(1)~Hz \cite{das-09}, and 0.368~Hz \cite{mitroy-08}, but
is three times more accurate.
%**********************************

\section{Hyperfine constants for $^{43}$Ca$^+$}

Calculations of hyperfine constants are carried out using the SD and SDpT all-order methods described in Section~\ref{E1}. A number of terms other
than terms $Z^{(a)}$ and $Z^{(c)}$ give significant contributions to the hyperfine constants. Therefore, scaling procedure described in Section
~\ref{E1} is not expected to produce more accurate values and is not carried out for the hyperfine constants.
 In
Table~\ref{tab-hyp}, we list hyperfine constants $A$ for $^{43}$Ca$^+$
 and compare our values  with available theoretical \cite{hyp-04} and experimental data \cite{hyp-expt-90,hyp-expt-98}.

In this table, we present the lowest-order $A^\text{DF}$, all-order $A^\text{SD}$, and $A^\text{SDpT}$ values for the $ns$, $np$, $nd$, and $nf$
levels up to  $n$ = 7. The magnetic moment  of $^{43}$Ca$^+$ used here ($I=7/2$, $\mu=-1.31727$)  is taken from \cite{web}. Our SDpT results are in
very good agreement with experimental results for the $ns$ and $np_{1/2}$ states when experimental uncertainties are taken into account. The
contributions from valence triple excitations are large for the hyperfine constants and have to be included for an accurate calculation.

Hyperfine constants $B$ (in MHz)  in
  $^{43}$Ca$^+$  are given in
  Table~\ref{tab-hypb}.
 Nuclear quadrupole moment $Q$ is taken to be
 equal to -0.044(9)~barns (1~b=10$^{-24}$cm$^2$) \cite{hyp-04}.
 The SD and SDpT  values are compared with theory \cite{hyp-04} and experiment
 ~\cite{hyp-expt-98}.
\section{Hyperfine-induced transition
polarizability of the $^{43}$Ca$^+$ ground state}

We now  turn to the calculation of the quadratic Stark shift  of the ground-state hyperfine interval ($F=4 - F=3$) in $^{43}$Ca$^+$. The quadratic
Stark shift is closely related to the blackbody radiation shift in the microwave frequency standards discussed, for example, in
Refs.~\cite{bbr-urn,safr-li-08,new-rb}. Our calculation follows the methodology outlined in those works.

The dominant second-order contribution to the polarizability cancels for the transition between the two hyperfine components of the $4s$ state.
Therefore,  the Stark shift of the hyperfine interval is governed by the
 the third-order $F$-dependent polarizability
$\alpha_{F}^{(3)}(0)$. The expression for the $\alpha_{F}^{(3)}(0)$ is ~\cite{bbr-urn}:
\begin{eqnarray}
\label{plrz} \alpha^{(3)}_F(0) & = & \frac{1}{3}\sqrt{(2I)(2I+1)(2I+2)}\left\{
\begin{array}{lll}
j_v &   I   & F \\
  I   &  j_v & 1
\end{array}
\right\} \times \nonumber \\  && g_I \mu_n \left(-1\right)^{F+I+j_v}\left(2T+C+R\right), \label{eq-bbr1}
\end{eqnarray}
where $g_I$ is the nuclear gyromagnetic ratio, $\mu_n$ is the nuclear magneton equal to 0.3924658 in $^{43}$Ca$^+$, $I=7/2$ is the nuclear spin, and
$j_v=1/2$ is the total angular momentum of the atomic ground state. The formulas for the $F$-independent terms $T$, $C$, and $R$  are given  in
Ref.~\cite{bbr-urn}. These terms are similar to the polarizability sum-over-state expression but are more complicated.

First, we calculate these values  in the DF approximation (in atomic units):
\begin{eqnarray}\label{eq-df}
2T^{\mathrm{DF}} &=&2.0018\times 10^{-4},\quad C^{\mathrm{DF}}=3.9507\times
10^{-7},\quad    \nonumber\\
R^{\mathrm{DF}} &=&3.8838\times 10^{-4}.
\end{eqnarray}

Since the  value of $C^{\rm DF}$ is smaller than the $T^{\rm DF}$ and $R^{\rm DF}$   by  three orders of magnitude, we do not recalculate the $C$
term  using the all-order method.

The expression for $R$  is  similar to that for $\alpha^{E1}$
 but contains
 diagonal hyperfine matrix element:
\[
\langle 4s\|  \mathcal{T}  \| 4s \rangle ^\text{SDpT}=  3.9629\times 10^{-7}\ \text{ a.u.}
\]

We use  our all-order recommended values for the reduced electric-dipole matrix elements  described in Section~\ref{E1} and their uncertainties to
calculate the main terms in the $T$ and $R$ sums. We refer to these values as the ``best set'' values. Available recommended NIST energies
\cite{nist-web} are used for $nl=4s-10s, 4p-6p$,  and SD energies are used for the other states up to $n=26$. The sum of $R$ terms with $n\leq 26$ is
equal to $R = 3.772(34)\times 10^{-4}$.
 The remainder of
the $R$ sum is evaluated in the DF approximation, $R_{n> 26} = 3.0\times 10^{-8}$, and  is less than 0.01\%.

\begin{table}
\caption{ Contributions to the $mp$ sums of term $2T$, $m=4-26$.  The main contribution $\sum_{n=5}^{26}$ calculated in the DF approximation is given
in the column labeled ``Main$^{\rm{DF}}$'' in $10^4$~a.u. The final values of the main contributions to  the $mp$ sums are given in the column
labeled ``Main$^{\rm{final}}$'' in $10^4$~a.u. Accumulated values are given for both DF and final results.  The ratio of the final and DF values for
the main terms is given in the fourth column in \%. The relative tail contribution $\sum_{n=27}^{70}$ calculated in the DF approximation is given in
the last column. \label{termT} }
\begin{ruledtabular}
\begin{tabular}{ccccc}
\multicolumn{1}{c}{$mp$}&
 \multicolumn{1}{c}{Main$^{\rm{DF}}$}&
 \multicolumn{1}{c}{Main$^{\rm{final}}$}&
 \multicolumn{1}{c}{Dif.(\%)} &
\multicolumn{1}{c}{Tail  (\%)}
 \\
 \hline
  $4p$ &    1.965&    2.060(12)&  4.6&     1.8\\
  $5p$ &    1.965&    2.079(12)&  5.5&     1.8\\
  $6p$ &    1.969&    2.089(12)&  5.7&     1.6\\
  $7p$ &    1.971&    2.094(12)&  5.9&     1.5\\
  $8p$ &    1.972&    2.097(12)&  5.9&     1.5\\
 $9pp$ &    1.973&    2.098(11)&  6.0&     1.4\\
 $10p$ &    1.974&    2.100(12)&  6.0&     1.4\\
 $11p$ &    1.974&    2.101(12)&  6.0&     1.4\\
 $12p$ &    1.974&    2.102(13)&  6.1&     1.4\\
 $13p$ &    1.975&    2.102(13)&  6.1&     1.4\\
 $14p$ &    1.975&    2.102(13)&  6.1&     1.4\\
 $15p$ &    1.975&    2.102(13)&  6.1&     1.4\\
 $16p$ &    1.975&    2.102(13)&  6.1&     1.4\\
 $17p$ &    1.975&    2.103(13)&  6.1&     1.3\\
 $18p$ &    1.976&    2.104(13)&  6.1&     1.3\\
 $19p$ &    1.976&    2.104(13)&  6.1&     1.3\\
 $20p$ &    1.976&    2.104(13)&  6.1&     1.3\\
 $21p$ &    1.979&    2.109(13)&  6.2&     1.2\\
 $22p$ &    1.979&    2.109(13)&  6.2&     1.2\\
 $23p$ &    1.983&    2.118(13)&  6.4&     0.9\\
 $24p$ &    1.986&    2.124(13)&  6.5&     0.8\\
 $25p$ &    1.988&    2.129(13)&  6.6&     0.7\\
 $26p$ &    1.990&    2.129(13)&  6.5&     0.6\\
 \end{tabular}
\end{ruledtabular}
\end{table}

%****************************
Term T contains two sums, over $ns$ and over $mp_{j}$. We evaluate main contributions, that include $n \leq 26$ and $m \leq 26$ using all-order
matrix elements and NIST or all-order energies as described above. We find that the remaining contributions with $n > 26$ and $m > 26$, are very
small. Table~\ref{termT} illustrates the size of the remainders and accuracy of the DF approximation. We break down each $mp$ term as
$$\sum_{mp} \left( \sum_{5s}^{26s} [...] + \sum_{27s}^{70s}[...]\right) $$
and
 list contributions to the $mp$ sums of term T, $m=4-26$ in Table~\ref{termT}.
Each $mp$ term is given by
\begin{equation}
\sum_{n=5}^{26} A_T  \frac{\langle 4s \|D \| mp_j\rangle \langle mp_j \|D \| ns\rangle \langle ns \| \mathcal{T}\| 4s\rangle}{\left( E_{mp} -
E_{4s}\right)\left( E_{ns} - E_{4s}\right)},
\end{equation}
where $A_T$ is an angular factor.
 The main contribution $\sum_{n=5}^{26}[...]$ calculated
in the DF approximation is given in the column labeled ``Main$^{\rm{DF}}$'' in $10^4$~a.u. The final values of the main contributions to  the $mp$
sums are given in the column labeled ``Main$^{\rm{final}}$'' in $10^4$~a.u. Accumulated values are given for both DF and final results to illustrate
the convergence of the $mp$ sum. The ratio of the final and DF values for the main terms is given in the fourth column in \%. The relative tail
contribution $\sum_{n=27}^{70}[...]$ calculated in the  DF approximation is given in the last column. The remainder is 0.6\% of the main term and is
equal to $0.013\times 10^{-4}$~a.u. Our final value for this term is 2$T^{\rm final} = 2.142(13)\times 10^{-4}$. Combining these contributions, we
obtain
\begin{equation}\label{eq-sum}
2T^{\text{final}}+C^{\text{DF}}+R^{\text{final}}=5.918(36)\times 10^{-4}\ \text{a.u.}
\end{equation}

The F-dependent factor in Eq.~(\ref{eq-bbr1}) is equal to  0.4609 for $F$ = 3 and -0.3585 for $F$ = 4. Using these values and the result from
Eq.~(\ref{eq-sum}), we obtain
\[
\alpha _{\mathrm{hf}}(4s) = \alpha _{F=4}^{(3)}(0)-\alpha _{F=3}^{(3)}(0)= -4.850(29)\times 10^{-4} \ \text{a.u.}
\]
The Stark shift coefficient $k$ defined as $\Delta \nu =kE^{2}$ is $ k=-\frac{1}{2}\left[ \alpha
_{F=4}^{(3)}(0)-\alpha_{F=3}^{(3)}(0)\right]$. Converting from atomic units, we obtain
$$
k=-2.425(15)\times 10^{-4}~\text{a.u} ~\text{\textrm{ = 6.03(4)}}\times 10^{-12}\ \text{Hz/(V/m)}^{2}.
$$
We note that the lowest-order DF value is $k^{\rm (DF)}$ = 6.00$\times 10^{-12}~\text{Hz/(V/m)}^{2}$. While values of both $R$ and $T$ terms change
with the inclusion of the correlation correction, it essentially cancels when these terms are added.

 The relative blackbody
radiative shift $\beta$ is defined as
\begin{equation}\label{eqres5}
\beta =-\frac{2}{15}\frac{1}{\nu _{\mathrm{hf}}}\left( \alpha \pi \right) ^{3}T^{4}\alpha _{\mathrm{hf}}(4s_{1/2})
\end{equation}
where ${\nu _{\mathrm{hf}}}$ is the $^{43}$Ca$^+$ hyperfine ($F = 3 - F = 4$) splitting equal to 3225.6082864(3)~MHz~\cite{hyp-94Z} and $T$ is
temperature taken to be  300~K. Using those factors and our value of $\alpha _{\mathrm{hf}}(4s)$ , we  obtain
\begin{equation*}
\beta =-2.6696\times 10^{-12}\alpha _{\mathrm{hf}}(4s) = 1.29(1)\times 10^{-15}.
\end{equation*}

\section{Conclusion}
A systematic study of Ca$^+$ atomic properties is carried out using high-precision  relativistic all-order
 method where all single, double, and partial triple excitations
of the Dirac-Fock wave function are included to all orders of perturbation theory. Energies, E1, E2, E3, matrix elements, transition rates,
lifetimes, $A$ and $B$ hyperfine constants, E1, E2, and E3 ground state polarizabilities, scalar E1 polarizabilities of the $5s$, $6s$, $7s$, $8s$,
$4p_j$, $5p_j$, $3d_j$, $4d_j$ states, and tensor polarizabilities of the  $4p_{3/2}$, $5p_{3/2}$, $3d_j$, and $4d_j$ states are calculated. We
evaluate the uncertainties of our calculations for most of the values listed in this work. The blackbody radiation (BBR) shift of the $4s - 3d_{5/2}$
clock transition in Ca$^+$
 is calculated to be $0.381(4)$~Hz at room temperature, $T=300$~K improving its accuracy by a factor of 3.
The quadratic Stark effect on hyperfine structure levels of $^{43}$Ca$^+$ ground state is investigated.  These calculations provide recommended
values critically evaluated for their accuracy for a number of Ca$^+$ atomic properties useful
for a variety of applications.\\

\begin{acknowledgments}
 The work of M.S.S.
was supported in part by National Science Foundation Grant  No.\
PHY-07-58088.
\end{acknowledgments}

%\bibliography{ca2}

\end{document}